%% file: use_cases_main.tex
\def\review{0} 
\def\paperversion{1} 
\def\arxivdisclaimer{1} 

\documentclass[10pt, conference, letterpaper]{IEEEtran}

\usepackage[noadjust]{cite}
\usepackage{amsmath,amssymb,amsfonts}
\usepackage{textcomp}
\usepackage{mathtools}
\usepackage{pgfplots}
\usepgfplotslibrary{groupplots}
\usepackage{adjustbox}
\usepackage{balance}
\usepackage{graphics}
\usepackage{bm}
\def\BibTeX{{\rm B\kern-.05em{\sc i\kern-.025em b}\kern-.08em
    T\kern-.1667em\lower.7ex\hbox{E}\kern-.125emX}}

\usepackage{supertabular}    
\usepackage{array} 
\include{./Material/colormap_inue_jetlight}

\if\review1
\usepackage{todonotes}
\else
\usepackage[disable]{todonotes}
\fi
\usepackage{numprint}     
\usepackage{makecell}   
\usepackage{xcolor,colortbl}
\usepackage{enumitem}            
\usepackage{longtable} 
\usepackage{booktabs}
\usepackage{makecell}
\usepackage[acronym]{glossaries}

\if\arxivdisclaimer0
\else
\usepackage{hyperref}
\glsdisablehyper
\fi
\usepackage{cleveref}

\usepackage{amsthm}
\usepackage{amsmath, amssymb}
\usepackage{tabu}

\usepackage{textcase}
\usepackage[tablename=TABLE,font=small]{caption}
\DeclareCaptionTextFormat{up}{\MakeTextUppercase{#1}}
\newtheorem{remark}{Remark}
\newtheorem*{remark*}{Remark}

\usepackage[utf8]{inputenc}

\let\oldtabular\tabular
\renewcommand{\tabular}{\small\oldtabular}

\usepackage{stackengine}

\newcommand{\egc}{e.\,g., }
\newcommand{\iec}{i.\,e., }

\newcommand{\wrt}{w.\,r.\,t.\ }

\newcolumntype{?}{!{\vrule width 1pt}}

\definecolor{mittelblau}{RGB}{0, 126, 198}
\definecolor{violettblau}{cmyk}{0.9, 0.6, 0, 0}
\definecolor{rot}{RGB}{238, 28 35}
\definecolor{apfelgruen}{RGB}{140, 198, 62}
\definecolor{gelb}{RGB}{1, 221, 0}
\definecolor{orange}{RGB}{244, 111, 33}
\definecolor{pink}{RGB}{237, 0, 140}
\definecolor{lila}{RGB}{128, 10, 145}
\definecolor{hellgrau}{RGB}{224, 224, 224}
\definecolor{mittelgrau}{RGB}{128, 128, 128}
\definecolor{dunkelgrau}{RGB}{80,80,80}
\definecolor{anthrazit}{RGB}{19, 31, 31}
\definecolor{darkgreen}{RGB}{0.125,0.5,0.169}

\if\arxivdisclaimer1
\else

\usepackage{balance}
\usepackage{lastpage}

\makeatletter
    \def\balanceissued{unbalanced}
    \let\oldbibitem\bibitem
    \def\bibitem{%
        \ifnum\thepage=\lastpage@lastpage%
            \expandafter\ifx\expandafter\relax\balanceissued\relax\else%
                \balance%
                \gdef\balanceissued{\relax}\fi%
            \else\fi%
        \oldbibitem}
\makeatother
\fi

\newcommand\blfootnote[1]{%
  \begingroup
  \renewcommand\thefootnote{}\footnote{#1}%
  \addtocounter{footnote}{-1}%
  \endgroup
}

\begin{document}

\title{Survey on Integrated Sensing and Communication Performance Modeling and Use Cases Feasibility}

\author{
    \IEEEauthorblockN{
        Silvio Mandelli\IEEEauthorrefmark{1},
        Marcus Henninger\IEEEauthorrefmark{1}\IEEEauthorrefmark{2},
        \if\paperversion1
        Maximilian Bauhofer\IEEEauthorrefmark{2}, 
        \fi
        Thorsten Wild\IEEEauthorrefmark{1}}
\IEEEauthorblockA{
\IEEEauthorrefmark{1}Nokia Bell Labs Stuttgart, 70469 Stuttgart, Germany\if\paperversion1
, \\        \IEEEauthorrefmark{2}Institute of Telecommunications, University of Stuttgart, 70569 Stuttgart, Germany 
\fi
 \\
silvio.mandelli@nokia-bell-labs.com}
}

\maketitle

\input{Content/Acronyms}
\input{Content/Abstract}

\if\arxivdisclaimer1
\blfootnote{This work has been submitted to the IEEE for possible publication. Copyright may be transferred without notice, after which this version may no longer be accessible.}
\else
\fi

\input{Content/Introduction}
\input{Content/Model}
\input{Content/SystemParameters}

\input{Content/SimulationStudy}
\input{Content/Use_Cases}
\input{Content/Conclusion}
\input{Content/Appendices}
\input{Content/Acknowledgment}

\bibliographystyle{IEEEtran}
\bibliography{references.bib}

\end{document}

%% file: Material/colormap_inue_jetlight.tex
\pgfplotsset{
    colormap={jet_inue}{
        rgb=(1, 1, 1)
        rgb=(0.99804, 0.99804, 0.99905)
        rgb=(0.99609, 0.99609, 0.99813)
        rgb=(0.99413, 0.99413, 0.99725)
        rgb=(0.99217, 0.99217, 0.99639)
        rgb=(0.99022, 0.99022, 0.99557)
        rgb=(0.98826, 0.98826, 0.99477)
        rgb=(0.9863, 0.9863, 0.99401)
        rgb=(0.98434, 0.98434, 0.99327)
        rgb=(0.98239, 0.98239, 0.99257)
        rgb=(0.98043, 0.98043, 0.9919)
        rgb=(0.97847, 0.97847, 0.99125)
        rgb=(0.97652, 0.97652, 0.99064)
        rgb=(0.97456, 0.97456, 0.99006)
        rgb=(0.9726, 0.9726, 0.98951)
        rgb=(0.97065, 0.97065, 0.98899)
        rgb=(0.96869, 0.96869, 0.9885)
        rgb=(0.96673, 0.96673, 0.98804)
        rgb=(0.96477, 0.96477, 0.98762)
        rgb=(0.96282, 0.96282, 0.98722)
        rgb=(0.96086, 0.96086, 0.98685)
        rgb=(0.9589, 0.9589, 0.98652)
        rgb=(0.95695, 0.95695, 0.98621)
        rgb=(0.95499, 0.95499, 0.98593)
        rgb=(0.95303, 0.95303, 0.98569)
        rgb=(0.95108, 0.95108, 0.98548)
        rgb=(0.94912, 0.94912, 0.98529)
        rgb=(0.94716, 0.94716, 0.98514)
        rgb=(0.94521, 0.94521, 0.98502)
        rgb=(0.94325, 0.94325, 0.98493)
        rgb=(0.94129, 0.94129, 0.98486)
        rgb=(0.93933, 0.93933, 0.98483)
        rgb=(0.93738, 0.93738, 0.98483)
        rgb=(0.93542, 0.93542, 0.98486)
        rgb=(0.93346, 0.93346, 0.98493)
        rgb=(0.93151, 0.93151, 0.98502)
        rgb=(0.92955, 0.92955, 0.98514)
        rgb=(0.92759, 0.92759, 0.98529)
        rgb=(0.92564, 0.92564, 0.98548)
        rgb=(0.92368, 0.92368, 0.98569)
        rgb=(0.92172, 0.92172, 0.98593)
        rgb=(0.91977, 0.91977, 0.98621)
        rgb=(0.91781, 0.91781, 0.98652)
        rgb=(0.91585, 0.91585, 0.98685)
        rgb=(0.91389, 0.91389, 0.98722)
        rgb=(0.91194, 0.91194, 0.98762)
        rgb=(0.90998, 0.90998, 0.98804)
        rgb=(0.90802, 0.90802, 0.9885)
        rgb=(0.90607, 0.90607, 0.98899)
        rgb=(0.90411, 0.90411, 0.98951)
        rgb=(0.90215, 0.90215, 0.99006)
        rgb=(0.9002, 0.9002, 0.99064)
        rgb=(0.89824, 0.89824, 0.99125)
        rgb=(0.89628, 0.89628, 0.9919)
        rgb=(0.89432, 0.89432, 0.99257)
        rgb=(0.89237, 0.89237, 0.99327)
        rgb=(0.89041, 0.89041, 0.99401)
        rgb=(0.88845, 0.88845, 0.99477)
        rgb=(0.8865, 0.8865, 0.99557)
        rgb=(0.88454, 0.88454, 0.99639)
        rgb=(0.88258, 0.88258, 0.99725)
        rgb=(0.88063, 0.88063, 0.99813)
        rgb=(0.87867, 0.87867, 0.99905)
        rgb=(0.87671, 0.87671, 1)
        rgb=(0.87476, 0.87573, 1)
        rgb=(0.8728, 0.87479, 1)
        rgb=(0.87084, 0.87387, 1)
        rgb=(0.86888, 0.87298, 1)
        rgb=(0.86693, 0.87213, 1)
        rgb=(0.86497, 0.8713, 1)
        rgb=(0.86301, 0.87051, 1)
        rgb=(0.86106, 0.86974, 1)
        rgb=(0.8591, 0.86901, 1)
        rgb=(0.85714, 0.8683, 1)
        rgb=(0.85519, 0.86763, 1)
        rgb=(0.85323, 0.86699, 1)
        rgb=(0.85127, 0.86638, 1)
        rgb=(0.84932, 0.8658, 1)
        rgb=(0.84736, 0.86525, 1)
        rgb=(0.8454, 0.86473, 1)
        rgb=(0.84344, 0.86424, 1)
        rgb=(0.84149, 0.86378, 1)
        rgb=(0.83953, 0.86335, 1)
        rgb=(0.83757, 0.86295, 1)
        rgb=(0.83562, 0.86259, 1)
        rgb=(0.83366, 0.86225, 1)
        rgb=(0.8317, 0.86194, 1)
        rgb=(0.82975, 0.86167, 1)
        rgb=(0.82779, 0.86142, 1)
        rgb=(0.82583, 0.86121, 1)
        rgb=(0.82387, 0.86103, 1)
        rgb=(0.82192, 0.86087, 1)
        rgb=(0.81996, 0.86075, 1)
        rgb=(0.818, 0.86066, 1)
        rgb=(0.81605, 0.8606, 1)
        rgb=(0.81409, 0.86057, 1)
        rgb=(0.81213, 0.86057, 1)
        rgb=(0.81018, 0.8606, 1)
        rgb=(0.80822, 0.86066, 1)
        rgb=(0.80626, 0.86075, 1)
        rgb=(0.80431, 0.86087, 1)
        rgb=(0.80235, 0.86103, 1)
        rgb=(0.80039, 0.86121, 1)
        rgb=(0.79843, 0.86142, 1)
        rgb=(0.79648, 0.86167, 1)
        rgb=(0.79452, 0.86194, 1)
        rgb=(0.79256, 0.86225, 1)
        rgb=(0.79061, 0.86259, 1)
        rgb=(0.78865, 0.86295, 1)
        rgb=(0.78669, 0.86335, 1)
        rgb=(0.78474, 0.86378, 1)
        rgb=(0.78278, 0.86424, 1)
        rgb=(0.78082, 0.86473, 1)
        rgb=(0.77886, 0.86525, 1)
        rgb=(0.77691, 0.8658, 1)
        rgb=(0.77495, 0.86638, 1)
        rgb=(0.77299, 0.86699, 1)
        rgb=(0.77104, 0.86763, 1)
        rgb=(0.76908, 0.8683, 1)
        rgb=(0.76712, 0.86901, 1)
        rgb=(0.76517, 0.86974, 1)
        rgb=(0.76321, 0.87051, 1)
        rgb=(0.76125, 0.8713, 1)
        rgb=(0.7593, 0.87213, 1)
        rgb=(0.75734, 0.87298, 1)
        rgb=(0.75538, 0.87387, 1)
        rgb=(0.75342, 0.87479, 1)
        rgb=(0.75147, 0.87573, 1)
        rgb=(0.74951, 0.87671, 1)
        rgb=(0.74755, 0.87772, 1)
        rgb=(0.7456, 0.87876, 1)
        rgb=(0.74364, 0.87983, 1)
        rgb=(0.74168, 0.88093, 1)
        rgb=(0.73973, 0.88206, 1)
        rgb=(0.73777, 0.88323, 1)
        rgb=(0.73581, 0.88442, 1)
        rgb=(0.73386, 0.88564, 1)
        rgb=(0.7319, 0.88689, 1)
        rgb=(0.72994, 0.88818, 1)
        rgb=(0.72798, 0.88949, 1)
        rgb=(0.72603, 0.89084, 1)
        rgb=(0.72407, 0.89222, 1)
        rgb=(0.72211, 0.89362, 1)
        rgb=(0.72016, 0.89506, 1)
        rgb=(0.7182, 0.89653, 1)
        rgb=(0.71624, 0.89802, 1)
        rgb=(0.71429, 0.89955, 1)
        rgb=(0.71233, 0.90111, 1)
        rgb=(0.71037, 0.9027, 1)
        rgb=(0.70841, 0.90432, 1)
        rgb=(0.70646, 0.90597, 1)
        rgb=(0.7045, 0.90766, 1)
        rgb=(0.70254, 0.90937, 1)
        rgb=(0.70059, 0.91111, 1)
        rgb=(0.69863, 0.91289, 1)
        rgb=(0.69667, 0.91469, 1)
        rgb=(0.69472, 0.91652, 1)
        rgb=(0.69276, 0.91839, 1)
        rgb=(0.6908, 0.92028, 1)
        rgb=(0.68885, 0.92221, 1)
        rgb=(0.68689, 0.92417, 1)
        rgb=(0.68493, 0.92616, 1)
        rgb=(0.68297, 0.92817, 1)
        rgb=(0.68102, 0.93022, 1)
        rgb=(0.67906, 0.9323, 1)
        rgb=(0.6771, 0.93441, 1)
        rgb=(0.67515, 0.93655, 1)
        rgb=(0.67319, 0.93872, 1)
        rgb=(0.67123, 0.94092, 1)
        rgb=(0.66928, 0.94316, 1)
        rgb=(0.66732, 0.94542, 1)
        rgb=(0.66536, 0.94771, 1)
        rgb=(0.66341, 0.95004, 1)
        rgb=(0.66145, 0.95239, 1)
        rgb=(0.65949, 0.95478, 1)
        rgb=(0.65753, 0.95719, 1)
        rgb=(0.65558, 0.95964, 1)
        rgb=(0.65362, 0.96211, 1)
        rgb=(0.65166, 0.96462, 1)
        rgb=(0.64971, 0.96716, 1)
        rgb=(0.64775, 0.96973, 1)
        rgb=(0.64579, 0.97233, 1)
        rgb=(0.64384, 0.97496, 1)
        rgb=(0.64188, 0.97762, 1)
        rgb=(0.63992, 0.98031, 1)
        rgb=(0.63796, 0.98303, 1)
        rgb=(0.63601, 0.98578, 1)
        rgb=(0.63405, 0.98856, 1)
        rgb=(0.63209, 0.99138, 1)
        rgb=(0.63014, 0.99422, 1)
        rgb=(0.62818, 0.9971, 1)
        rgb=(0.62622, 1, 1)
        rgb=(0.6272, 1, 0.99706)
        rgb=(0.62821, 1, 0.9941)
        rgb=(0.62925, 1, 0.9911)
        rgb=(0.63032, 1, 0.98807)
        rgb=(0.63142, 1, 0.98502)
        rgb=(0.63255, 1, 0.98193)
        rgb=(0.63371, 1, 0.97881)
        rgb=(0.63491, 1, 0.97566)
        rgb=(0.63613, 1, 0.97248)
        rgb=(0.63738, 1, 0.96927)
        rgb=(0.63867, 1, 0.96603)
        rgb=(0.63998, 1, 0.96276)
        rgb=(0.64133, 1, 0.95945)
        rgb=(0.6427, 1, 0.95612)
        rgb=(0.64411, 1, 0.95276)
        rgb=(0.64555, 1, 0.94936)
        rgb=(0.64702, 1, 0.94594)
        rgb=(0.64851, 1, 0.94248)
        rgb=(0.65004, 1, 0.939)
        rgb=(0.6516, 1, 0.93548)
        rgb=(0.65319, 1, 0.93193)
        rgb=(0.65481, 1, 0.92836)
        rgb=(0.65646, 1, 0.92475)
        rgb=(0.65815, 1, 0.92111)
        rgb=(0.65986, 1, 0.91744)
        rgb=(0.6616, 1, 0.91374)
        rgb=(0.66337, 1, 0.91001)
        rgb=(0.66518, 1, 0.90625)
        rgb=(0.66701, 1, 0.90246)
        rgb=(0.66888, 1, 0.89864)
        rgb=(0.67077, 1, 0.89478)
        rgb=(0.6727, 1, 0.8909)
        rgb=(0.67466, 1, 0.88699)
        rgb=(0.67665, 1, 0.88304)
        rgb=(0.67866, 1, 0.87907)
        rgb=(0.68071, 1, 0.87506)
        rgb=(0.68279, 1, 0.87102)
        rgb=(0.6849, 1, 0.86696)
        rgb=(0.68704, 1, 0.86286)
        rgb=(0.68921, 1, 0.85873)
        rgb=(0.69141, 1, 0.85457)
        rgb=(0.69365, 1, 0.85039)
        rgb=(0.69591, 1, 0.84617)
        rgb=(0.6982, 1, 0.84192)
        rgb=(0.70053, 1, 0.83763)
        rgb=(0.70288, 1, 0.83332)
        rgb=(0.70527, 1, 0.82898)
        rgb=(0.70768, 1, 0.82461)
        rgb=(0.71013, 1, 0.82021)
        rgb=(0.7126, 1, 0.81577)
        rgb=(0.71511, 1, 0.81131)
        rgb=(0.71765, 1, 0.80681)
        rgb=(0.72022, 1, 0.80229)
        rgb=(0.72282, 1, 0.79773)
        rgb=(0.72545, 1, 0.79314)
        rgb=(0.72811, 1, 0.78853)
        rgb=(0.7308, 1, 0.78388)
        rgb=(0.73352, 1, 0.7792)
        rgb=(0.73627, 1, 0.77449)
        rgb=(0.73905, 1, 0.76975)
        rgb=(0.74187, 1, 0.76498)
        rgb=(0.74471, 1, 0.76018)
        rgb=(0.74758, 1, 0.75535)
        rgb=(0.75049, 1, 0.75049)
        rgb=(0.75342, 1, 0.7456)
        rgb=(0.75639, 1, 0.74067)
        rgb=(0.75939, 1, 0.73572)
        rgb=(0.76241, 1, 0.73074)
        rgb=(0.76547, 1, 0.72572)
        rgb=(0.76856, 1, 0.72068)
        rgb=(0.77168, 1, 0.7156)
        rgb=(0.77483, 1, 0.71049)
        rgb=(0.77801, 1, 0.70536)
        rgb=(0.78122, 1, 0.70019)
        rgb=(0.78446, 1, 0.69499)
        rgb=(0.78773, 1, 0.68976)
        rgb=(0.79103, 1, 0.6845)
        rgb=(0.79437, 1, 0.67921)
        rgb=(0.79773, 1, 0.67389)
        rgb=(0.80113, 1, 0.66854)
        rgb=(0.80455, 1, 0.66316)
        rgb=(0.80801, 1, 0.65775)
        rgb=(0.81149, 1, 0.65231)
        rgb=(0.81501, 1, 0.64683)
        rgb=(0.81855, 1, 0.64133)
        rgb=(0.82213, 1, 0.63579)
        rgb=(0.82574, 1, 0.63023)
        rgb=(0.82938, 1, 0.62463)
        rgb=(0.83305, 1, 0.61901)
        rgb=(0.83675, 1, 0.61335)
        rgb=(0.84048, 1, 0.60766)
        rgb=(0.84424, 1, 0.60194)
        rgb=(0.84803, 1, 0.5962)
        rgb=(0.85185, 1, 0.59042)
        rgb=(0.85571, 1, 0.58461)
        rgb=(0.85959, 1, 0.57877)
        rgb=(0.8635, 1, 0.5729)
        rgb=(0.86745, 1, 0.56699)
        rgb=(0.87142, 1, 0.56106)
        rgb=(0.87543, 1, 0.5551)
        rgb=(0.87946, 1, 0.54911)
        rgb=(0.88353, 1, 0.54308)
        rgb=(0.88763, 1, 0.53703)
        rgb=(0.89176, 1, 0.53094)
        rgb=(0.89591, 1, 0.52483)
        rgb=(0.9001, 1, 0.51868)
        rgb=(0.90432, 1, 0.51251)
        rgb=(0.90857, 1, 0.5063)
        rgb=(0.91285, 1, 0.50006)
        rgb=(0.91717, 1, 0.49379)
        rgb=(0.92151, 1, 0.48749)
        rgb=(0.92588, 1, 0.48116)
        rgb=(0.93028, 1, 0.4748)
        rgb=(0.93472, 1, 0.46841)
        rgb=(0.93918, 1, 0.46199)
        rgb=(0.94368, 1, 0.45554)
        rgb=(0.9482, 1, 0.44906)
        rgb=(0.95276, 1, 0.44255)
        rgb=(0.95734, 1, 0.436)
        rgb=(0.96196, 1, 0.42943)
        rgb=(0.96661, 1, 0.42282)
        rgb=(0.97129, 1, 0.41619)
        rgb=(0.976, 1, 0.40952)
        rgb=(0.98074, 1, 0.40283)
        rgb=(0.98551, 1, 0.3961)
        rgb=(0.99031, 1, 0.38934)
        rgb=(0.99514, 1, 0.38255)
        rgb=(1, 1, 0.37573)
        rgb=(1, 0.99511, 0.37378)
        rgb=(1, 0.99018, 0.37182)
        rgb=(1, 0.98523, 0.36986)
        rgb=(1, 0.98025, 0.36791)
        rgb=(1, 0.97523, 0.36595)
        rgb=(1, 0.97019, 0.36399)
        rgb=(1, 0.96511, 0.36204)
        rgb=(1, 0.96, 0.36008)
        rgb=(1, 0.95487, 0.35812)
        rgb=(1, 0.9497, 0.35616)
        rgb=(1, 0.9445, 0.35421)
        rgb=(1, 0.93927, 0.35225)
        rgb=(1, 0.93401, 0.35029)
        rgb=(1, 0.92872, 0.34834)
        rgb=(1, 0.9234, 0.34638)
        rgb=(1, 0.91805, 0.34442)
        rgb=(1, 0.91267, 0.34247)
        rgb=(1, 0.90726, 0.34051)
        rgb=(1, 0.90182, 0.33855)
        rgb=(1, 0.89634, 0.33659)
        rgb=(1, 0.89084, 0.33464)
        rgb=(1, 0.8853, 0.33268)
        rgb=(1, 0.87974, 0.33072)
        rgb=(1, 0.87414, 0.32877)
        rgb=(1, 0.86852, 0.32681)
        rgb=(1, 0.86286, 0.32485)
        rgb=(1, 0.85717, 0.3229)
        rgb=(1, 0.85146, 0.32094)
        rgb=(1, 0.84571, 0.31898)
        rgb=(1, 0.83993, 0.31703)
        rgb=(1, 0.83412, 0.31507)
        rgb=(1, 0.82828, 0.31311)
        rgb=(1, 0.82241, 0.31115)
        rgb=(1, 0.81651, 0.3092)
        rgb=(1, 0.81057, 0.30724)
        rgb=(1, 0.80461, 0.30528)
        rgb=(1, 0.79862, 0.30333)
        rgb=(1, 0.79259, 0.30137)
        rgb=(1, 0.78654, 0.29941)
        rgb=(1, 0.78045, 0.29746)
        rgb=(1, 0.77434, 0.2955)
        rgb=(1, 0.76819, 0.29354)
        rgb=(1, 0.76202, 0.29159)
        rgb=(1, 0.75581, 0.28963)
        rgb=(1, 0.74957, 0.28767)
        rgb=(1, 0.7433, 0.28571)
        rgb=(1, 0.737, 0.28376)
        rgb=(1, 0.73068, 0.2818)
        rgb=(1, 0.72432, 0.27984)
        rgb=(1, 0.71792, 0.27789)
        rgb=(1, 0.7115, 0.27593)
        rgb=(1, 0.70505, 0.27397)
        rgb=(1, 0.69857, 0.27202)
        rgb=(1, 0.69206, 0.27006)
        rgb=(1, 0.68551, 0.2681)
        rgb=(1, 0.67894, 0.26614)
        rgb=(1, 0.67233, 0.26419)
        rgb=(1, 0.6657, 0.26223)
        rgb=(1, 0.65903, 0.26027)
        rgb=(1, 0.65234, 0.25832)
        rgb=(1, 0.64561, 0.25636)
        rgb=(1, 0.63885, 0.2544)
        rgb=(1, 0.63206, 0.25245)
        rgb=(1, 0.62524, 0.25049)
        rgb=(1, 0.6184, 0.24853)
        rgb=(1, 0.61152, 0.24658)
        rgb=(1, 0.6046, 0.24462)
        rgb=(1, 0.59766, 0.24266)
        rgb=(1, 0.59069, 0.2407)
        rgb=(1, 0.58369, 0.23875)
        rgb=(1, 0.57666, 0.23679)
        rgb=(1, 0.56959, 0.23483)
        rgb=(1, 0.5625, 0.23288)
        rgb=(1, 0.55538, 0.23092)
        rgb=(1, 0.54822, 0.22896)
        rgb=(1, 0.54103, 0.22701)
        rgb=(1, 0.53382, 0.22505)
        rgb=(1, 0.52657, 0.22309)
        rgb=(1, 0.51929, 0.22114)
        rgb=(1, 0.51199, 0.21918)
        rgb=(1, 0.50465, 0.21722)
        rgb=(1, 0.49728, 0.21526)
        rgb=(1, 0.48988, 0.21331)
        rgb=(1, 0.48245, 0.21135)
        rgb=(1, 0.47499, 0.20939)
        rgb=(1, 0.4675, 0.20744)
        rgb=(1, 0.45997, 0.20548)
        rgb=(1, 0.45242, 0.20352)
        rgb=(1, 0.44484, 0.20157)
        rgb=(1, 0.43722, 0.19961)
        rgb=(1, 0.42958, 0.19765)
        rgb=(1, 0.42191, 0.19569)
        rgb=(1, 0.4142, 0.19374)
        rgb=(1, 0.40646, 0.19178)
        rgb=(1, 0.3987, 0.18982)
        rgb=(1, 0.3909, 0.18787)
        rgb=(1, 0.38307, 0.18591)
        rgb=(1, 0.37521, 0.18395)
        rgb=(1, 0.36733, 0.182)
        rgb=(1, 0.35941, 0.18004)
        rgb=(1, 0.35146, 0.17808)
        rgb=(1, 0.34347, 0.17613)
        rgb=(1, 0.33546, 0.17417)
        rgb=(1, 0.32742, 0.17221)
        rgb=(1, 0.31935, 0.17025)
        rgb=(1, 0.31125, 0.1683)
        rgb=(1, 0.30311, 0.16634)
        rgb=(1, 0.29495, 0.16438)
        rgb=(1, 0.28675, 0.16243)
        rgb=(1, 0.27853, 0.16047)
        rgb=(1, 0.27027, 0.15851)
        rgb=(1, 0.26199, 0.15656)
        rgb=(1, 0.25367, 0.1546)
        rgb=(1, 0.24532, 0.15264)
        rgb=(1, 0.23694, 0.15068)
        rgb=(1, 0.22853, 0.14873)
        rgb=(1, 0.2201, 0.14677)
        rgb=(1, 0.21163, 0.14481)
        rgb=(1, 0.20312, 0.14286)
        rgb=(1, 0.19459, 0.1409)
        rgb=(1, 0.18603, 0.13894)
        rgb=(1, 0.17744, 0.13699)
        rgb=(1, 0.16882, 0.13503)
        rgb=(1, 0.16016, 0.13307)
        rgb=(1, 0.15148, 0.13112)
        rgb=(1, 0.14277, 0.12916)
        rgb=(1, 0.13402, 0.1272)
        rgb=(1, 0.12524, 0.12524)
        rgb=(0.99315, 0.12329, 0.12329)
        rgb=(0.98627, 0.12133, 0.12133)
        rgb=(0.97936, 0.11937, 0.11937)
        rgb=(0.97242, 0.11742, 0.11742)
        rgb=(0.96545, 0.11546, 0.11546)
        rgb=(0.95845, 0.1135, 0.1135)
        rgb=(0.95141, 0.11155, 0.11155)
        rgb=(0.94435, 0.10959, 0.10959)
        rgb=(0.93726, 0.10763, 0.10763)
        rgb=(0.93013, 0.10568, 0.10568)
        rgb=(0.92298, 0.10372, 0.10372)
        rgb=(0.91579, 0.10176, 0.10176)
        rgb=(0.90857, 0.099804, 0.099804)
        rgb=(0.90133, 0.097847, 0.097847)
        rgb=(0.89405, 0.09589, 0.09589)
        rgb=(0.88674, 0.093933, 0.093933)
        rgb=(0.8794, 0.091977, 0.091977)
        rgb=(0.87203, 0.09002, 0.09002)
        rgb=(0.86463, 0.088063, 0.088063)
        rgb=(0.8572, 0.086106, 0.086106)
        rgb=(0.84974, 0.084149, 0.084149)
        rgb=(0.84225, 0.082192, 0.082192)
        rgb=(0.83473, 0.080235, 0.080235)
        rgb=(0.82718, 0.078278, 0.078278)
        rgb=(0.81959, 0.076321, 0.076321)
        rgb=(0.81198, 0.074364, 0.074364)
        rgb=(0.80434, 0.072407, 0.072407)
        rgb=(0.79666, 0.07045, 0.07045)
        rgb=(0.78896, 0.068493, 0.068493)
        rgb=(0.78122, 0.066536, 0.066536)
        rgb=(0.77345, 0.064579, 0.064579)
        rgb=(0.76566, 0.062622, 0.062622)
        rgb=(0.75783, 0.060665, 0.060665)
        rgb=(0.74997, 0.058708, 0.058708)
        rgb=(0.74208, 0.056751, 0.056751)
        rgb=(0.73416, 0.054795, 0.054795)
        rgb=(0.72621, 0.052838, 0.052838)
        rgb=(0.71823, 0.050881, 0.050881)
        rgb=(0.71022, 0.048924, 0.048924)
        rgb=(0.70218, 0.046967, 0.046967)
        rgb=(0.6941, 0.04501, 0.04501)
        rgb=(0.686, 0.043053, 0.043053)
        rgb=(0.67787, 0.041096, 0.041096)
        rgb=(0.6697, 0.039139, 0.039139)
        rgb=(0.66151, 0.037182, 0.037182)
        rgb=(0.65328, 0.035225, 0.035225)
        rgb=(0.64503, 0.033268, 0.033268)
        rgb=(0.63674, 0.031311, 0.031311)
        rgb=(0.62842, 0.029354, 0.029354)
        rgb=(0.62008, 0.027397, 0.027397)
        rgb=(0.6117, 0.02544, 0.02544)
        rgb=(0.60329, 0.023483, 0.023483)
        rgb=(0.59485, 0.021526, 0.021526)
        rgb=(0.58638, 0.019569, 0.019569)
        rgb=(0.57788, 0.017613, 0.017613)
        rgb=(0.56935, 0.015656, 0.015656)
        rgb=(0.56079, 0.013699, 0.013699)
        rgb=(0.5522, 0.011742, 0.011742)
        rgb=(0.54357, 0.0097847, 0.0097847)
        rgb=(0.53492, 0.0078278, 0.0078278)
        rgb=(0.52624, 0.0058708, 0.0058708)
        rgb=(0.51752, 0.0039139, 0.0039139)
        rgb=(0.50878, 0.0019569, 0.0019569)
        rgb=(0.5, 0, 0)
    }
}

%% file: Content/Acronyms.tex
\newacronym{5G}{5G}{fifth generation}
\newacronym{6G}{6G}{sixth generation}
\newacronym{adc}{ADC}{analog-to-digital-converter}
\newacronym{agv}{AGV}{automated guided vehicle}
\newacronym{crlb}{CRLB}{Cram\'er-Rao lower bound}
\newacronym{dl}{DL}{downlink}
\newacronym{fft}{FFT}{fast Fourier transform}
\newacronym{fr1}{FR1}{Frequency Range 1}
\newacronym{fr2}{FR2}{Frequency Range 2}
\newacronym{fr3}{FR3}{Frequency Range 3}
\newacronym{isac}{ISAC}{Integrated Sensing and Communication}
\newacronym[plural=KPIs,firstplural = key performance indicators (KPIs)]{kpi}{KPI}{key performance indicator}
\newacronym{los}{LoS}{line of sight}
\newacronym[plural=NAFs,firstplural = normalized angular frequencies (NAFs)]{naf}{NAF}{normalized angular frequency}
\newacronym{mn}{MN}{Mobile Networks}
\newacronym{nlos}{NLOS}{non-line of sight}
\newacronym{ofdm}{OFDM}{orthogonal frequency division multiplexing}
\newacronym{papr}{PAPR}{peak-to-average power ratio}
\newacronym{poc}{PoC}{Proof of Concept}
\newacronym{prs}{PRS}{positioning reference signal}
\newacronym{rcs}{RCS}{radar cross section}
\newacronym{snr}{SNR}{signal-to-noise ratio}
\newacronym{sqnr}{SQNR}{signal-to-quantization-noise ratio}
\newacronym{tdd}{TDD}{time division duplex}
\newacronym{trp}{TRP}{transmission and reception point}
\newacronym{ue}{UE}{user equipment}
\newacronym{ula}{ULA}{uniform linear array}
\newacronym{ura}{URA}{uniform rectangular array}

%% file: Content/Abstract.tex
\begin{abstract}
\if\paperversion1 
As the research community starts to address the key features of 6G cellular standards, one of the agreed bridge topics to be studied already in 5G advanced releases is \gls{isac}. 
The first efforts of the research community are focusing on \gls{isac} enablers, fundamental limits, and first demonstrators, that show that the time has come for the deployment of sensing functionalities in cellular standards.

This survey paper takes a needed step towards \gls{isac} deployment, providing an analytical toolkit to model cellular systems' sensing performance, accounting for both their fundamental and practical constraints. We then elaborate on the likely features of 6G systems to provide the feasible sensing \glspl{kpi} in the frequency ranges spanned by cellular networks, including the potential new bands available in 6G, the \gls{fr3}. 

We further validate our framework by visually investigating \gls{isac} constraints with simulation examples. Finally, we assess the feasibility of few selected scenarios that can be enabled by \gls{isac}, highlighting in each of them the limiting factor and, thus, which gaps should be filled by the research and standardization communities in the next years. 
\else
We introduce a framework to validate sensing use cases \wrt \glspl{kpi} such as maximum achievable sensing range, achievable accuracy, and achievable resolution. We then focus on currently available Nokia products in \gls{fr1} and \gls{fr2} as well as a projected \gls{fr3} system parameterization to estimate the performance in some specific sensing scenarios of interest.
\fi
\end{abstract}
\glsresetall
\begin{IEEEkeywords}
Integrated sensing and communication (ISAC), 6G, performance modeling, use cases feasibility. 
\end{IEEEkeywords}

%% file: Content/Introduction.tex
\section{Introduction}\label{sec:Introduction}
The next generations of wireless networks will define the key enablers for the applications and services of the next decade. Focusing on cellular systems, the \gls{6G} cellular standard aims at boosting communications performance and flexibility compared to the previous \gls{5G} releases~\cite{viswanathan2020communications}. Among the disruptive topics studied by the \gls{6G} research community, \gls{isac} has been raising interest, given the revolution that it brings in communication systems' hardware, software, and architecture~\cite{liu2022integrated,wild2023integrated}.
In particular, the increased carrier bandwidths enabled at higher frequencies together with their aggregation opportunities~\cite{3gpp_38301_1,3gpp_38301_2}, and the deployment of massive antenna arrays~\cite{liu2022integrated}, made possible to consider running wireless networks as radars. This is a considerable step from \gls{5G}'s active localization, where the goal was to locate active users as in~\cite{henninger2022probabilistic}, that were collaborating in their localization. With \gls{6G} \gls{isac}, the wireless network scans and acquires information from a passive environment.

Accordingly, the \gls{isac} topic has already been object of different conferences and workshops,  recapping the basics in comprehensive surveys~\cite{liu2022integrated,liu2022survey} or public projects' deliverables~\cite{hexa_x}. The authors of this paper have participated in the world's first \gls{isac} demonstrator at mmWave (28 GHz), performing both communications and mono-static sensing~\cite{wild2023integrated}, showcased at the Mobile World Congress 2023 in Barcelona.

To the best of our knowledge, the existing works on \gls{isac} focus on optimization of its performance and enablers, e.g., on the necessary hardware and architecture design~\cite{wild2023integrated}. The existing surveys~\cite{liu2022integrated,liu2022survey} discuss use cases and relevant \glspl{kpi}.
On the other hand, the radar literature~\cite{braun2014ofdm} neglects the constraints and parameterization that is needed when also communications has to be considered.
The closest work to ours~\cite{liu2022survey} provides an excellent analytical study on the sensing \glspl{kpi}, focusing on achievable performance bounds in the presence of noise, and the effect of resolution on performance. However, the authors do not provide considerations on the inter-relationships between the effects that may limit sensing performance in \gls{isac} and do not comment on the performance achievable, according to \gls{6G} systems' foreseen features and parameters.
Therefore, in this survey we aim at jointly analyzing the relevant effects that determine the sensing \glspl{kpi} of an \gls{isac} system, focusing on typical radar impairments and the limitations given by communications features. In particular, this paper provides:
\begin{itemize}
    \item The needed analytical tools to assess the impact of thermal noise, quantization noise, and resolution in a single \gls{isac} system (Section \ref{sec:model}).
    \item In Section~\ref{sec:systemParameters}, a characterization of feasible \gls{isac} systems' features and parameters for each of the different frequency ranges that are currently being considered for \gls{6G}~\cite{viswanathan2020communications} or are already in use for 5G, except the sub-THz spectrum which we expect to initially not play the strongest role in early 6G ~\cite{wild2023integrated}:
    \begin{itemize}
        \item \gls{fr1}: from 600 MHz to 6 GHz,
        \item \gls{fr2}: mmWave, 24 GHz to 71 GHz,
        \item the new \gls{fr3}, that is not yet specified: from 7 to 20 GHz.
    \end{itemize}
    \item Visual examples of the different constraints to be considered in \gls{isac} based on simulation data in Section~\ref{sec:SimulationStudy}, guiding the reader through the relevant limits to be considered.
    \item An evaluation of selected use cases in Section~\ref{sec:use_cases}, based on the analytical model provided in this work, determining their feasibility and performance according to the foreseen features of \gls{6G} systems. This Section will highlight what are the main limiting factors constraining \gls{isac} \glspl{kpi}, defining the most pertinent questions to be addressed in \gls{isac} cellular systems' research.
    \item A conclusion together with a summary of the main findings in Section~\ref{sec:conclusion}.
\end{itemize}



%% file: Content/Model.tex
\section{Assumed Model}\label{sec:model}
We first lay down the model used to analytically evaluate scenarios. In this document, we are considering a mono-static setup with quasi co-located \glspl{ura} TX and RX of equal characteristics, with $R$ rows and $C$ columns, spaced $\Delta r$ and $ \Delta c$, respectively. This practically coincides with a full duplex sensing system, that is the likely implementation of first \gls{isac} deployments~\cite{wild2023integrated}. Note that equal performance, apart from angular resolution, could be achieved with a bi-static deployment. Even though additional gains may be achieved by fusing information between multiple points at a central function, this is not considered in this work. 
The symbols and notations used throughout this document are defined in Table~\ref{tab:Notation} at the end of this document. 

\subsection{Link budget with thermal noise}
\label{subsec:LinkBudget}
In order to get meaningful information, like most of the radar literature~\cite{braun2014ofdm}, we assume line of sight propagation to write the received power equation as function of the range
\begin{align}
    P_R(r) &= \frac{P_T G_T}{4\pi r^2}\Psi \frac{1}{4 \pi r^2}\frac{G_R \lambda^2}{4\pi} =
\nonumber \\
&= {P_T G_T G_R} \cdot \Psi \cdot \frac{c_0^2}{(4\pi)^3 r^4 f_c^2} \;.
\label{eq:RxPower}
\end{align}
Note that in \cite{braun2014ofdm}, the transmit antenna $G_T$ is not included (\iec isotropic radiation is assumed), but in this document we consider it to account for the TX beamforming gains, assuming that we are perfectly focusing the target with the used transmit beam. 
The noise power can be written as
\begin{equation}
    P_N = (N_0 F) (N \Delta f) \;,
\end{equation}
which allows us to define the available \gls{snr} on each single \gls{ofdm} symbol 
\begin{equation}
    \gamma_S (r) = \frac{P_R(r)}{P_N} \;.
    \label{eq:snrBeforeProcessingGains}
\end{equation}
The \gls{snr} is increased due to a multiplicative gain given by the number of subcarriers $N$ and \gls{ofdm} symbols $M$~\cite{braun2014ofdm}, resulting in an \gls{snr} of
\begin{equation}
    \gamma (r) = \frac{P_R(r)}{P_N} NM \;,
    \label{eq:snrAfterProcessingGains}
\end{equation}
which in case of the periodogram is due to the focusing of the Fourier transform operations on the sparse representation of the radar echo. 

To compute the achievable range in each scenario, an \gls{snr} of $\gamma^* = 17$~dB (after antenna gain, processing gain, etc.) is necessary to achieve robust performance, i.e., matching the \gls{crlb} on accuracy to attain reliable performance in terms of both false alarms and missed detections~\cite{MUSIC_collapse}. Then, for each scenario the maximum achievable range due to the noise limit can be computed as
\begin{align}
r_n^* &: \gamma(r_n^*) = \gamma^* =  17 \text{ dB} \approx 50 \; \Rightarrow 
\\
r_n^* &= \sqrt[4]{\frac{P_T G_T G_R \Psi c_0^2 N M}{\gamma^*(4\pi)^3 f_c^2 N_0 F N \Delta f}} =
\nonumber \\
 &= \sqrt[4]{\frac{P_T G_T G_R \Psi c_0^2 M}{\gamma^*(4\pi)^3 f_c^2 N_0 F \Delta f}} \; .
\label{eq:AchievableRangeNoise}
\end{align}

\if\paperversion1
\else
\begin{remark}
Note that ``fade margins'' could be modeled by considering that the \gls{rcs} $\Psi$ is a random variable, depending on the object's orientation. In this way, one could establish the needed margin, choosing a value for $\Psi$ corresponding to the distribution's quantile $\epsilon = 1 - \chi$, where $\chi$ is the desired reliability.
\end{remark}
\fi

\subsection{Quantization noise}
However, the achievable range may in practice be dictated by the strongest reflection of an object (or the direct path as self-interference). This is due to the quantization noise of a $Q$ bit \gls{adc}. One should determine the impact of the objects present in the environment, grouped in the set $\mathcal{T}$. Moreover, also self-interference between transmitter and receiver must be considered. Accordingly, the limiting factor is given by
\begin{equation}
\overline{t} = \text{max} \left[ \text{max}_{t \in \mathcal{T}} \left(\frac{\Psi_t    }{r_t^4} \right), \frac{\alpha 4 \pi} { r_{t'}^{2}} \right] \;,
\label{eq:StrongestPath}
\end{equation}
where $\Psi_t$ and $r_t$ are the $t$-th object's \gls{rcs} and range. 
The isolation between transmitter and receiver $\alpha \leq 1$ depends on beamforming and hardware features, of transmitter and receiver, separated by $r_{t'}$~\cite{nwankwo2017survey}.
Similarly, one could trivially rewrite the self-interference part of~\eqref{eq:StrongestPath}, in case of co-located transmitter and receiver.
Accordingly, the \gls{sqnr} can be written as~\cite{bennett1948spectra}
\begin{align}
\text{\gls{sqnr}}_Q = (2^Q)^2 \approx 6.02 \cdot Q\; \text{dB} \; .
\label{eq:sqnr}
\end{align}
Taking distortions due to the \gls{papr} of the transmitted \gls{ofdm} signal into account as $\gamma_{\text{PAPR}}$, which can \egc be modeled as in \cite{behravan2002papr}, and assuming perfect automatic gain control, the \gls{sqnr} at the receiver is
\begin{align}
\gamma_Q = \frac{\text{\gls{sqnr}}_Q MN}{\gamma_{\text{PAPR}}}  \; ,
\label{eq:sqnr1}
\end{align}
where the $MN$ processing gain is still available after \gls{adc} quantization. Additional losses in terms of non-ideal automatic gain control can further reduce $\gamma_Q$.
Note that in case of quantization with $Q'$ bits in the \gls{fft} operations used for sensing, the processing gain and \gls{ofdm}'s \gls{papr} should not be considered, leading to ${\gamma_q}(Q, Q') = \min (\gamma_Q, \text{\gls{sqnr}}_{Q'})$. This is because both \gls{fft} inputs and outputs are quantized, thus leading to a quantization of the complex numbers, whose amplitude squared is already the periodogram.
Accordingly, the maximum achievable range $r^*_q$ (guaranteeing an \gls{snr} of $\gamma^*$) due to quantization noise of a target with \gls{rcs} $\Psi$ is given by 
\begin{align}
\frac{\Psi}{\overline{t} \left( r_q^*\right)^4}  &=  \frac{\gamma^*}{{\gamma_q}(Q, Q')} \; \Rightarrow 
\nonumber \\
r_q^* &= \sqrt[4]{\frac{\Psi \gamma_q(Q, Q') }{\overline{t} {\gamma^*}}}
\; .
\label{eq:AchievableRangeQuant}
\end{align}
In case of being limited by an object at range $r_{\overline{t}}$ with \gls{rcs} $\Psi_{\overline{t}}$, the achievable range can be written as
\begin{equation}
r_q^* = r_{\overline{t}} \cdot \sqrt[4]{\frac{\Psi \gamma_q(Q, Q') }{\Psi_{\overline{t}} {\gamma^*}}}\;.
\end{equation}

\subsection{Achievable accuracy}

We recall that super-resolution techniques - like MUSIC~\cite{MUSIC_collapse} -- and interpolation-based algorithms -- like SARA~\cite{mandelli2022sampling} - achieve the \gls{crlb} with $\gamma(r) \geq \gamma^*$. This means that the standard deviation of sensing estimates in this operating regime, \iec $r \leq r^*$, is given by the already derived \gls{crlb} formulae in~\cite{braun2014ofdm}, that are
\begin{align}
\sigma_r &= \frac{c}{4\pi \Delta f} \sqrt{\frac{6}{(N^2-1)\gamma(r)}}  \; ,
\label{eq:AccuracyRange} \\
\sigma_s &= \frac{c}{4\pi f_c T_O} \sqrt{\frac{6}{(M^2-1)\gamma(r)}}  \; ,
\label{eq:AccuracySpeed} \\
\sigma_z &=  \frac{1}{2\pi } \sqrt{\frac{6}{(R^2-1)\gamma(r)}} \; ,
\label{eq:AccuracyElevation} \\
\sigma_x &=  \frac{1}{2\pi } \sqrt{\frac{6}{(C^2-1)\gamma(r)}} \; 
\label{eq:AccuracyAzimuth} \; .
\end{align}
\begin{remark}
In case of clock errors between transmitter and receiver, with standard deviations of absolute clock and frequency errors -- after correction algorithms -- $\sigma_t$ and $\sigma_f$, respectively, the range and speed standard deviations can be updated as
\begin{align}
\sigma'_r &= \sqrt{\sigma_r^2 + (c\sigma_t)^2}  \; ,
 \\
\sigma'_s &= \sqrt{\sigma_s^2 + \left(\frac{c}{f_c} \right)^2 \sigma_f^2}  \; ,
\end{align}
where we assumed independency between clock error and thermal noise.
\end{remark}
For angular measures, $\sigma_x$ and $\sigma_z$ are the corresponding \glspl{naf} in horizontal (x-axis) and vertical (z-axis) direction of the receive array, respectively, described in~\cite{mandelli2022sampling}.
The mapping of $\sigma_x$ and $\sigma_z$ to angular accuracy depends on the incident angle with respect to the system, with the highest performance at boresight.
Once the incident azimuth $\theta$ and elevation $\phi$ are known, their \gls{naf} can be computed as
\begin{align}
\eta &= \frac{\Delta r}{\lambda}\sin (\phi) \; ,\\
\ell &= \frac{\Delta c}{\lambda}\frac{\sin (\theta)}{ \cos (\phi)} \; .
\end{align}
Then, one can use inversion formulae to measure the offset corresponding to $\sigma_z$ and $\sigma_x$ in angles, as follows
\begin{align}
\sigma_\phi &= \sin^{-1} \left( \frac{\lambda}{\Delta r} \left(\eta + \frac{\sigma_z}{2}\right) \right) - \sin^{-1} \left( \frac{\lambda}{\Delta r} \left(\eta - \frac{\sigma_z}{2}\right) \right) \; , \\
\sigma_\theta &= \sin^{-1} \left( \frac{\lambda}{\Delta c }\cos(\phi) \left(\ell + \frac{\sigma_x}{2}\right) \right) \nonumber \\
&- \sin^{-1} \left( \frac{\lambda}{\Delta c }\cos(\phi) \left(\ell 
- \frac{\sigma_x}{2}\right) \right)\; ,
\end{align}
where we account for the offset asymmetry in positive and negative directions for angles deviating from boresight.

\subsection{Achievable Resolution}

For the achievable resolutions, see definitions in Table~\ref{tab:Notation}, the equations from~\cite{braun2014ofdm, hoctor1990unifying} can be used to get 
\begin{align}
\rho_r &= \frac{c}{2 N \Delta f }  \; ,
\label{eq:ResolutionRange} \\
\rho_s &= \frac{c}{2 T_f f_c}  \; ,
\label{eq:ResolutionSpeed} \\
\rho_z &=  \frac{1}{2R - 1} \; , 
\label{eq:ResolutionZ} \\
\rho_x &= \frac{1}{2C - 1} \; ,
\label{eq:ResolutionX} 
\end{align}
where $\rho_z$ and $\rho_x$ are determined using the shape, \iec vertical and horizontal, respectively, of the sum co-array of transmit and receive array~\cite{hoctor1990unifying}. 
We recall that the sum co-array of two \glspl{ura} with $R \times C$ elements is given by a \gls{ura} with $(2R-1) \times (2C-1)$ elements.

The shift to resolutions $\rho_{\phi}$ and $\rho_{\theta}$ in terms of the incident angles $\phi$ and $\theta$ is obtained in a way similar to the accuracy as 
\begin{align}
\rho_{\phi} &= \sin^{-1} \left( \frac{\lambda}{\Delta r} \left(\eta + \frac{\rho_z}{2} \right) \right) - \sin^{-1} \left( \frac{\lambda}{\Delta r} \left(\eta - \frac{\rho_z}{2} \right) \right)  \; ,
\label{eq:ResolutionElevation} \\ 
\rho_{\theta} &= \sin^{-1} \left( \frac{\lambda}{\Delta c } \cos(\phi) \left(\ell + \frac{\rho_x}{2} \right) \right) \nonumber \\
&- \sin^{-1} \left( \frac{\lambda}{\Delta c } \cos(\phi) \left(\ell - \frac{\rho_x}{2} \right) \right) \; .
\label{eq:ResolutionAzimuth}  
\end{align}
These angular resolutions can be further translated into a required spacing in meters between two objects at distance $r$ from the base station. For simplicity, we consider the vertical direction resolution $\rho_v$ (depending only on the elevation $\phi$) and the horizontal direction resolution $\rho_h$ (depending on both azimuth $\theta$ and elevation $\phi$) separately. They are given as
\begin{align}
\rho_v &= r\cdot \sin \left( \rho_{\phi} \right) \; ,
\label{eq:VerticalResolution} \\ 
\rho_h  &= r\cdot \sin \left( \rho_{\theta} \right) \; .
\label{eq:HorizontalResolution}  
\end{align}
This allows us also to write the maximum range achievable due to the angular resolution $r_a^*$, if we require a vertical/horizontal resolution of $\rho_v^*,\rho_h^*$, respectively, as
\begin{align}
r_v^* &= \frac{\rho_v^*}{\sin (\rho_\phi)} \; ,
\\
r_h^* &= \frac{\rho_h^*}{\sin (\rho_\theta)} \; .
\end{align}
However, targets need to be separated just in one domain, thus the best performing resolution direction dominates. Therefore, one should consider the maximum between $r_h^*$ and $r_v^*$, and only if range and speed resolutions do not allow separating  targets reliably.

\subsection{Unambiguous Ranges}

The achievable performance may further be limited by the unambiguous ranges (\iec maximum values without ambiguities due to aliasing). For range and speed, the maximum unambiguous values are 
\begin{align}
r_u^* &= \frac{c_0}{2\Delta f} \; ,
\label{eq:UnambRange} \\ 
s_u &= \frac{c_0}{2 f_c T_0} \; .
\label{eq:UnambSpeed}
\end{align}
\subsection{Achievable sensing range}
All the considerations in the previous subsections allow us to finally write the achievable sensing range $r^*$ as the most stringent constraint imposed by thermal noise, quantization noise, angular resolution, and unambiguous range. In particular, we have  
\begin{equation}
r^* = \min \left( r_n^*, r_q^*, \max(r_v^*, r_h^*), r_u^* \right) \;,
\label{eq:AchievableRangeAll}
\end{equation}
if range and speed resolution do not allow to reliably separate targets in the use case of interest. Otherwise, \egc in case of moving target detection with a high enough speed resolution to separate the target from static clutter, one could neglect angular resolution dependencies as follows
\begin{equation}
r^* = \min \left( r_n^*, r_q^*, r_u^* \right) \;.
\label{eq:AchievableRangeNoResolution}
\end{equation}
\if\paperversion1
\begin{remark}
\Cref{eq:AchievableRangeAll,eq:AchievableRangeNoResolution} assume \gls{los} between object and radar, which may not be the case in practice. \gls{los} probability is, however, hard to model, and therefore left out of this discussion. 
\end{remark}
\else
\begin{remark}
\Cref{eq:AchievableRangeAll,eq:AchievableRangeNoResolution} assume \gls{los} between object and radar, which may not be the case in practice. \gls{los} probability, however, depends heavily on the scenario, making it difficult to draw general assumptions. Nokia internal raytracing data evaluated in an urban scenario (Minneapolis downtown) revealed that ca. 60\% of \gls{ue} positions were in \gls{los}, which gives a first indication at the order of magnitude that can be expected for the \gls{los} probability in such scenarios.
\end{remark}
\fi

%% file: Content/SystemParameters.tex
\section{System Parameters}\label{sec:systemParameters}
\if\paperversion1
\input{Content/SystemParameters_Publication}
\else
\input{Content/SystemParameters_Internal}
\fi

%% file: Content/SystemParameters_Publication.tex
In this Section, we illustrate the likely parameters of the next generation of cellular networks. The proposed values for \gls{fr1}, \gls{fr2}, and \gls{fr3} are listed in Table~\ref{tab:sys_params}, and they will be used in Section~\ref{sec:use_cases} to evaluate use cases' feasibility. In every frequency range, we assume antenna radiator elements with gain $G_E = 2 = 3$ dBi. The resulting array gain accounts also for the number of elements, as follows

\begin{equation}
G_T = RCG_E \;.
\end{equation}
For \gls{fr1}, the values are based on currently available white papers by the industry~\cite{wesemann2023energy,holma2021extreme}. Note that the $24 \times 8$ elements are given by assuming 6 radiators vertically stacked per antenna port, resulting in an equivalent $4 \times 8$ digital system. 
The radio transmission features are taken from Table 5.3.2-1 of TS 38.101-1~\cite{3gpp_38301_1}, dictating the number of subcarriers $N$ given the specific carrier aggregation parameters. In \gls{fr1}, we assume   
$2 \times 100$~MHz carrier aggregation, corresponding to $N = 6552$ and transmit power for outdoor scenarios $P_{T,O} = 49$ dBm.

For \gls{fr2}, we based our assumptions on a running mmWave system at 28 GHz, showcased at 2023 Mobile World Congress~\cite{wild2023integrated}. We doubled the array size to $32 \times 32$ and the total bandwidth to $8 \times 200$~MHz,  accounting for the evolution of hardware in the next years.

For \gls{fr3}, we assume 1024 elements, as most of the recent literature~\cite{wesemann2023energy}, doubling the bandwidth of \gls{fr1} systems to $4 \times 100$~MHz, using the same transmit power $P_{T,O} = 49$ dBm as in \gls{fr1} for outdoor scenarios.

Considering indoor use cases, the European regulations~\cite{recommendation1999limitation} 
on electromagnetic field exposure fix  the power density limit to $S_0 = 10$ W/m\textsuperscript{2} in the frequencies of interest. Accordingly, assuming a minimum distance of $d' = 1$ m between transmitter and humans, the transmit power limit for indoor scenarios is
\begin{equation}
    P_{T,I} = \frac{S_0 4 \pi d'^2}{G_T T^* P^*} \; ,
\end{equation}
where $T^*$ and $P^*$ are the \gls{tdd} duty-cycle and the power reduction factor applicable due to beam steering and time averaging, respectively. In this work, we consider $T^* = 80\%$, while $P^*=0.25$ is assumed from~\cite{baracca2018statistical}.

To determine the number of \gls{ofdm} symbols $M$, we consider the \gls{prs} for sensing in the \gls{dl}, whose configuration options are defined in~\cite{3gpp_38211}. Configuring the number of \gls{ofdm} symbols that are allocated for \gls{prs} per slot $L_{\text{PRS}}$ and the comb size $K^{\text{PRS}}_{\text{comb}}$ such that the number of transmitted \gls{prs} symbols per slot is maximized leads to $M^{\text{PRS}}_{\text{slot}}=6$ (option $\{L_{\text{PRS}} = 12, K^{\text{PRS}}_{\text{comb}} = 2\}$ \cite{3gpp_38211}). Then, choosing offsets and periodicity properly, \gls{prs} can be configured for transmission in every slot. The final values for the number of \gls{ofdm} symbols per radio frame $M$ are obtained by accounting for a $1-T^* = 20\%$ \gls{tdd} overhead.
\begin{table}[hbt!]%
\centering
\caption{Considered System Parameterizations}
\label{tab:sys_params}
\begin{tabular}{|c?c|c|c|}
\hline
\textbf{Parameter} & \textbf{\gls{fr1}} & \textbf{\gls{fr2}} & \textbf{\gls{fr3}}   \\ 
\Xhline{2.5\arrayrulewidth}
$f_c$ & 3.5 GHz & 28 GHz  & 7 GHz\\ \hline
$B$ & 200 MHz & 1600 MHz & 400 MHz \\ \hline
$\Delta f$ & 30 kHz & 120 kHz & 60 kHz \\ \hline
$T_0$ & 35.67 \textmu s & 8.92 \textmu s & 17.84 \textmu s  \\ \hline
$N$ & 6552 &  12672 & 6480 \\ \hline
$M$ & 96 & 384 & 192 \\ \hline
$F$ & 8 dB & 8 dB & 8 dB \\ \hline
$G_T$ & 25.8 dB & 33 dB & 33 dB   \\ \hline
$G_E$ & 3 dBi & 3 dBi & 3 dBi \\ \hline
$R, C$ & 24, 8 & 32, 32 & 32, 32 \\ \hline
$\Delta r, \Delta c$ & 0.7$\lambda$, 0.5$\lambda$ &  0.5$\lambda$, 0.5$\lambda$ & 0.5$\lambda$, 0.5$\lambda$ \\ \hline
$P_{T,O}$ & 49 dBm & 36 dBm & 49 dBm  \\ \hline
$P_{T,I}$ & 32.2 dBm & 25 dBm & 25 dBm  \\ \hline
$Q$ & 12 & 12 & 12 \\ \hline
\end{tabular}
\end{table}

%% file: Content/SimulationStudy.tex
\section{Visual examples of \gls{isac} limits}\label{sec:SimulationStudy}

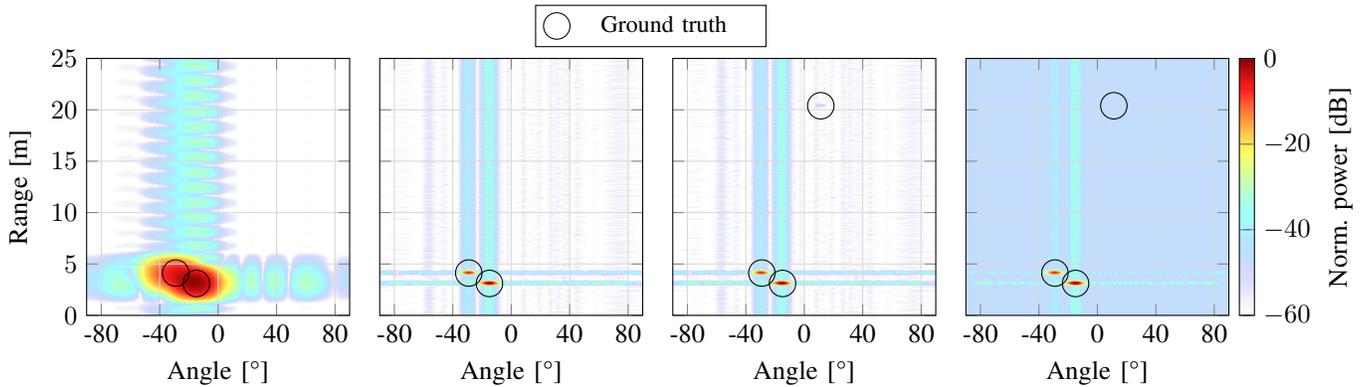
\begin{figure*}[t]
	\centering
    \input{Content/periodograms.tikz}
    \caption{Periodograms of exemplary scene including thermal noise and configuration according to Tab.~\ref{tab:sys_params}: (1) one observable target for FR1, (2) higher resolution in FR2 enables separation of two close targets, (3) third target further away distinguishable, (4) third target covered by quantization noise after $Q'=8\,$bit quantization.}
	\label{fig:periodogram_examples}
\end{figure*}

Before discussing selected use cases in Section~\ref{sec:use_cases}, we provide visualizations of some of the discussed \gls{isac} constraints introduced in Section~\ref{sec:model}.
Those examples are intended to give the reader a better understanding in order to make the following use cases discussion easier to follow.
\\In Fig.~\ref{fig:periodogram_examples}, four different range-angle periodograms are visualized. The two periodograms on the left illustrate the effect of resolution. The leftmost image is obtained with the \gls{fr1} system parameterization from Table~\ref{tab:sys_params}, with two close targets in both range and angle. One can observe that the poor resolution due to the limited antenna aperture and available bandwidth does not allow discriminating the two close targets (marked by black circles), resulting in the detection of only a single one. In the periodogram right beside the first, on the other hand, both targets are resolvable due to the \gls{fr2} system providing higher bandwidth and array horizontal aperture, resulting in sufficient range and angle (\iec spatial) resolution capabilities. 
\\ We continue by discussing the impact of quantization noise. The second periodogram from the right displays the case where no quantization noise is considered, enabling the detection of an additional target that is further away at a range of $\approx 20\,$m. In the rightmost figure, however, a further quantization in the \gls{fft} operations is assumed with $Q'=8\,$bits. This causes the farther target to ``drown" in quantization noise, blinding it out from the resulting periodogram and rendering its detection infeasible.
\\These simplified examples serve to show that resolution capabilities and quantization might play a critical role in determining the achievable performance of \gls{isac} systems. Therefore, they should be taken into account for the following use cases investigation, together with the link budget considerations of Subsection~\ref{subsec:LinkBudget}.

%% file: Content/periodograms.tikz
\begin{tikzpicture}
    \centering
    \begin{groupplot}[
        group style={group size=4 by 1, vertical sep=0mm, horizontal sep=4mm},
        width=0.28\textwidth, height=5cm,
        grid=major, grid style={solid,gray!30},
        xmin=-pi/2, xmax=pi/2, xlabel near ticks,
        xtick={-1.3963,-0.6981,0,0.6981,1.3963},
        xticklabels={-80,-40,0,40,80},
        xlabel near ticks,
        ymin=0, ymax=25, ylabel near ticks,
        ytick={0,5,...,25},
        ylabel near ticks,
        point meta min=-60,
        point meta max=0,
        colormap name=jet_inue,
		colorbar style={ylabel={Norm. power [dB]}, at={(1.04,0)}, anchor=south west, width=2mm}, 
		legend style={fill=white, fill opacity=0.4, draw opacity=1, text opacity=1, nodes={scale=1, transform shape}, at={(1,1)}, anchor=east, /tikz/every even column/.append style={column sep=0.1cm},
        font=\small,
        {minimum width=2.5 cm}},
		legend cell align={left},
		legend columns=4,
		view={0}{90},
        axis on top, ]

        \nextgroupplot[ylabel={Range [m]}, ylabel style = {yshift=-0mm}, xlabel={Angle [\textdegree]}, xlabel style = {xshift=0mm}, legend to name = gt]
            \addlegendimage{only marks, mark=o, mark size=5pt}
            \addlegendentry{Ground truth}
            \addplot graphics [xmin=-1.57, xmax=1.57, ymin=0, ymax=25]{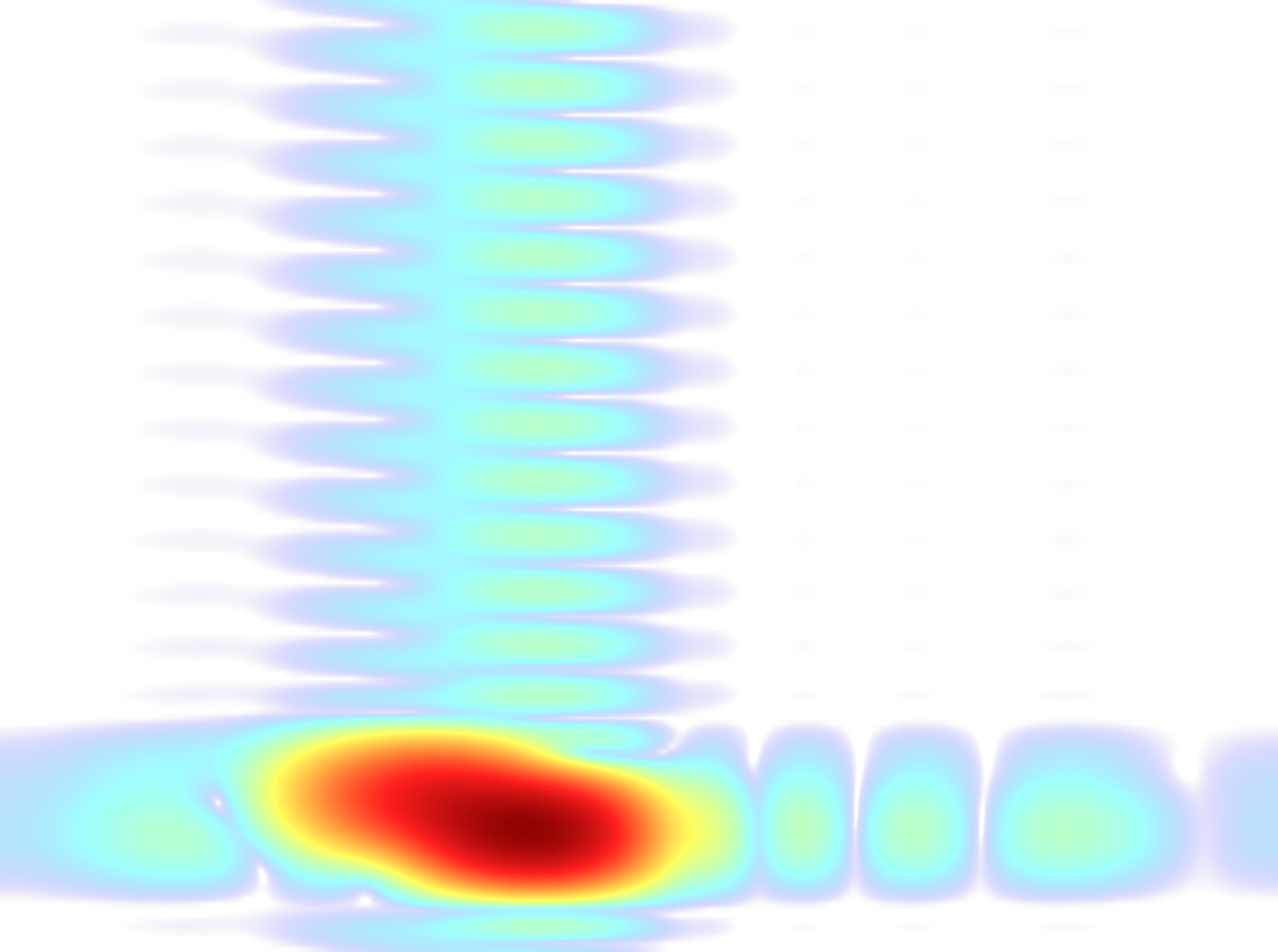};
            \addplot[mark=o,only marks,mark size=5pt,color=black, forget plot] table[x=angle,y=distance] {
                angle distance
                -0.507 4.118
                -0.260 3.105
                };

        \nextgroupplot[yticklabels=\empty, xlabel={Angle [\textdegree]}, xlabel style = {xshift=0mm}, ]
            \addplot graphics [xmin=-1.57, xmax=1.57, ymin=0, ymax=25]{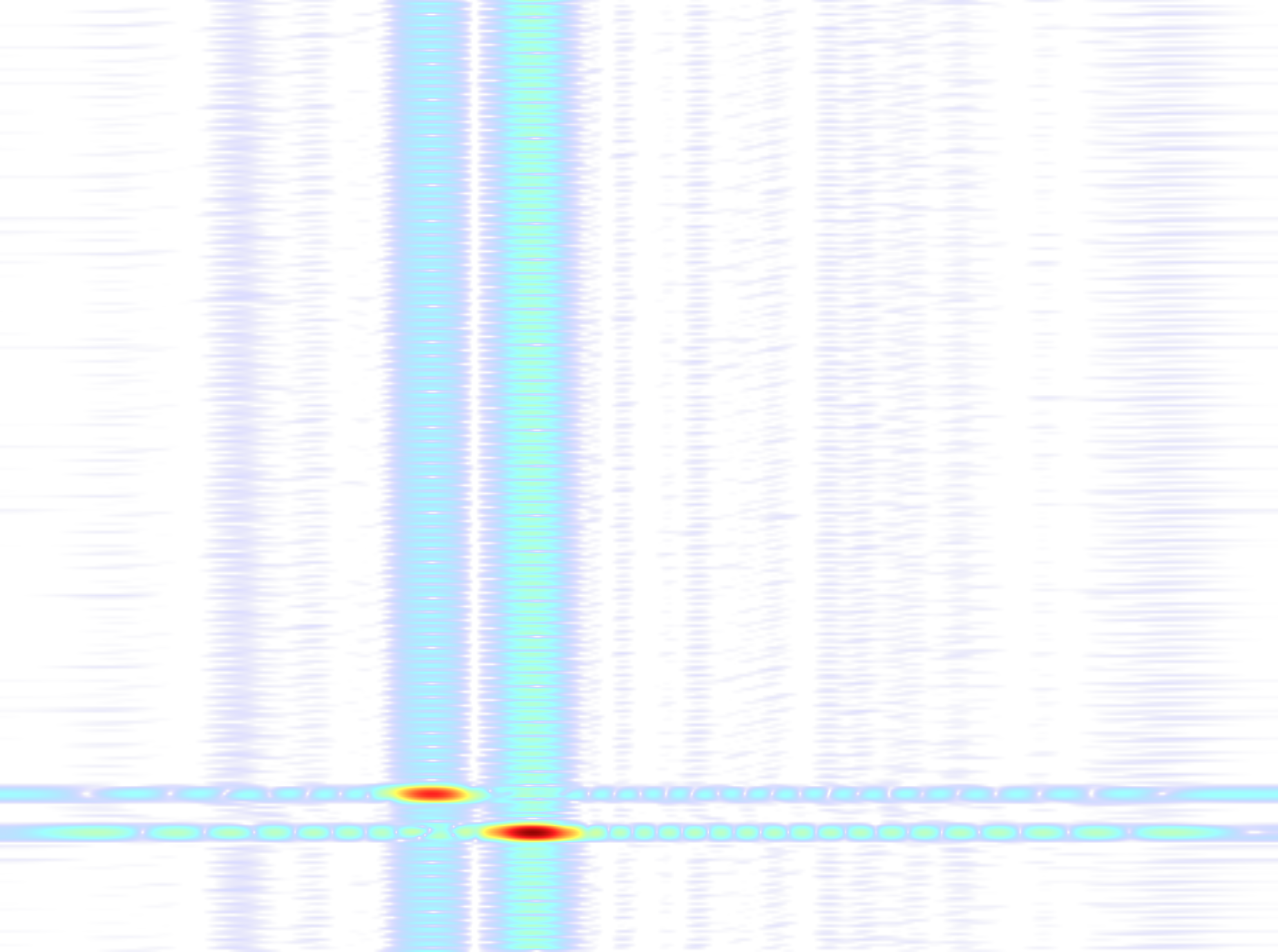};
            \addplot[forget plot, mark=o,only marks,mark size=5pt,color=black, forget plot] table[x=angle,y=distance] {
                angle distance
                -0.507 4.118
                -0.260 3.105
                };

        \nextgroupplot[yticklabels=\empty, xlabel={Angle [\textdegree]}, xlabel style = {xshift=0mm}, ]
            \addplot graphics [xmin=-1.57, xmax=1.57, ymin=0, ymax=25]{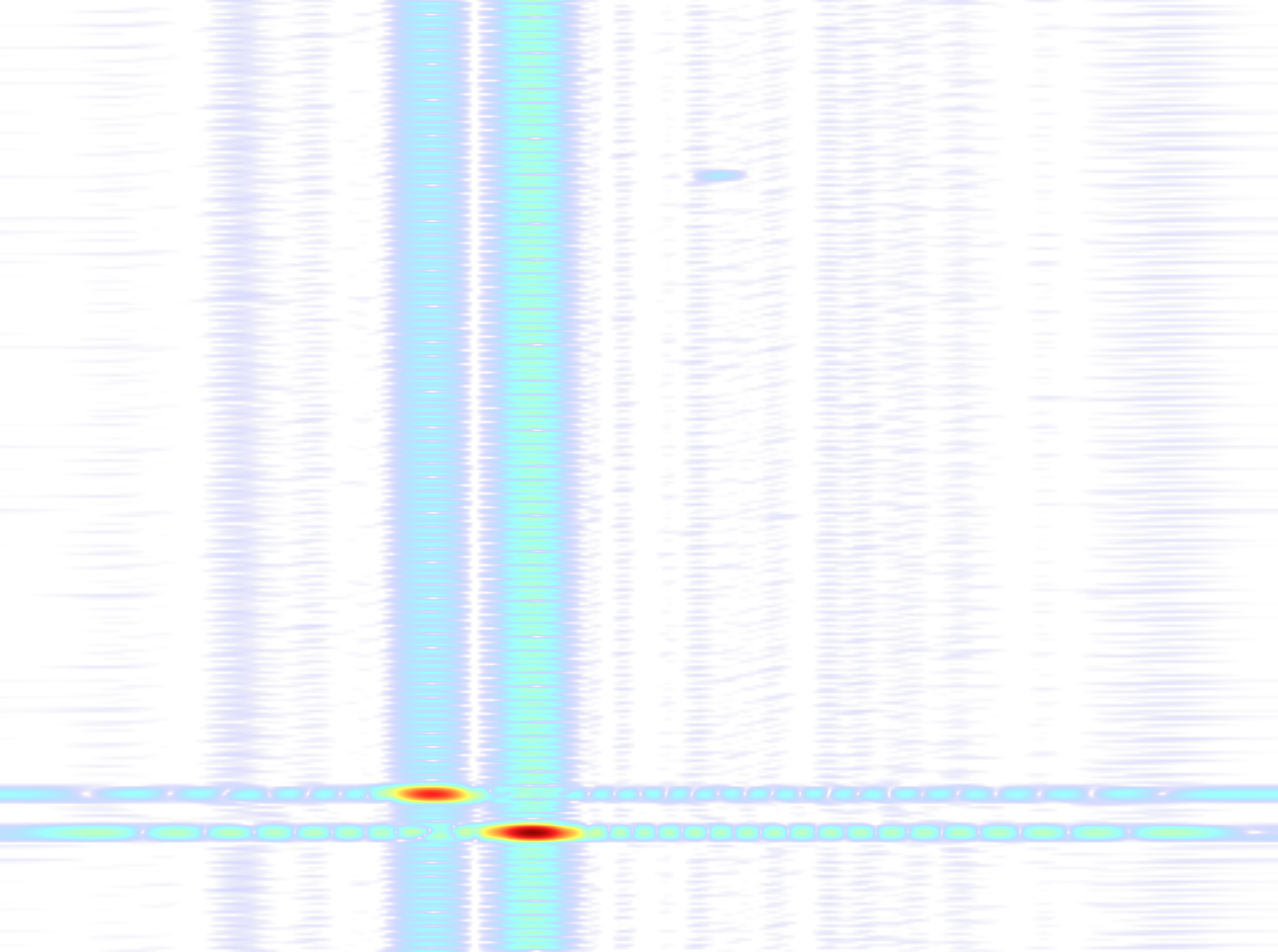};
            \addplot[forget plot, mark=o,only marks,mark size=5pt,color=black, forget plot] table[x=angle,y=distance] {
                angle distance
                -0.507 4.118
                -0.260 3.105
                0.197 20.396
                };

        \nextgroupplot[yticklabels=\empty, colorbar, xlabel={Angle [\textdegree]}, xlabel style = {xshift=0mm}, ]
            \addplot graphics [xmin=-1.57, xmax=1.57, ymin=0, ymax=25]{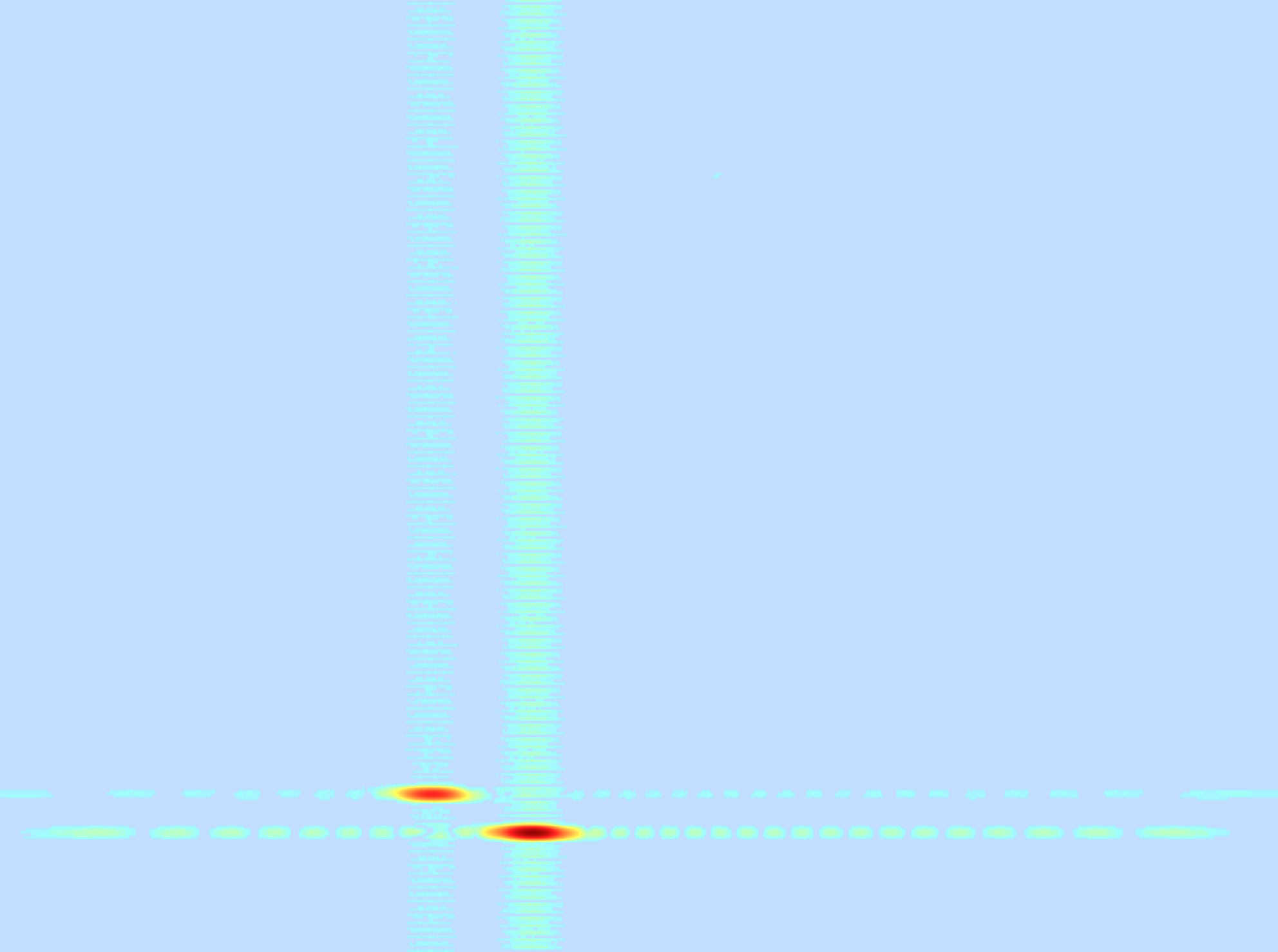};
            \addplot[forget plot, mark=o,only marks,mark size=5pt,color=black, forget plot] table[x=angle,y=distance] {
                angle distance
                -0.507 4.118
                -0.260 3.105
                0.197 20.396
                };

	\end{groupplot}    
    \node at (7.5, 3.85) {\pgfplotslegendfromname{gt}};
\end{tikzpicture}

%% file: Content/Use_Cases.tex
\if\paperversion1
\input{Content/Use_Cases_Publication}
\else
\input{Content/Use_Cases_Internal}
\fi

%% file: Content/Use_Cases_Publication.tex
\section{Use Cases Investigation}\label{sec:use_cases}

In this section, we investigate three families of emerging \gls{isac} use cases: 1) Indoor Factory Safety (\ref{sec:indoor_fac}), 2) Roadway Monitoring (\ref{sec:highway_mon}), and 3) Outside Drone Detection (\ref{sec:drone_detection}). 
\\First, assuming \gls{los} and using the foundations established in the previous sections, Table~\ref{tab:fr1_fr2_perf_params} displays a list of achievable sensing \gls{kpi}s in \gls{fr1} and \gls{fr2}, at the range at which the minimum required sensing \gls{snr} of $\gamma^* = 17$~dB is attained.
For each use case, we will focus on the most relevant \glspl{kpi} and from those we try to draw conclusions on the scenario's feasibility.

\begin{table}[hbt!]
\centering
\caption{Sensing Performance Parameters with SNR = 17 \MakeTextLowercase{d}B) and L\MakeTextLowercase{o}S condition}
\label{tab:fr1_fr2_perf_params}
\begin{tabular}{|l?c|c|c|}
\hline
 \textbf{Parameter} & \textbf{\gls{fr1}} & \textbf{\gls{fr2}} & \textbf{\gls{fr3}}  \\ 
\Xhline{2.5\arrayrulewidth}
$\sigma_r$ [m] & 0.042 & 0.005 & 0.021 \\ \hline
$\sigma_v$ [m/s] & 0.689 & 0.086 & 0.345 \\ \hline
$\sigma_{\phi}$ (boresight) [$^{\circ}$]  & 0.188 & 0.197 & 0.197 \\ \hline
$\sigma_{\theta}$ (boresight)  [$^{\circ}$] & 0.795 & 0.197 & 0.197 \\ \hline
$\rho_r$ [m]  & 0.76 & 0.1 & 0.39 \\ \hline
$\rho_s$ [m/s] & 4.29 & 0.54 & 2.14 \\ \hline
$\rho_{\phi}$ (boresight)  [$^{\circ}$] & 1.74 & 1.82 & 1.82 \\ \hline
$\rho_{\theta}$ (boresight) [$^{\circ}$] & 7.64 & 1.82 & 1.82 \\ \hline
$\rho_v$ (boresight)  [m] & $ 0.03r$ & $0.032r$ & $0.032r$ \\ \hline
$\rho_h$ (boresight) [m] & $ 0.133r$ & $0.032r$ & $0.032r$ \\ \hline
$r_u^*$ [m] & 5000 & 1250 & 2500 \\ \hline 
$s_u$ [m/s] & 600.7 & 300.3 & 600.6 \\ \hline
\end{tabular}
\end{table}


Furthermore, Table~\ref{tab:max_sense_ranges} lists the maximum achievable sensing ranges due to the noise limit (Eq.~\eqref{eq:AccuracyRange}) for objects that are of interest for the investigated use cases, along with their assumed \gls{rcs} values from literature~\cite{skolnik1980introduction, de2019drone}. As one can deduce from Table~\ref{tab:max_sense_ranges}, sensing performance is -- for the systems we assumed in Section~\ref{sec:systemParameters} -- not limited by thermal noise.
In addition to resorting to literature, an alternative way to estimate the \gls{rcs} is suggested by~\cite{braun2014ofdm}. One could take measurements with a given setup, estimate the periodogram's peak value $\hat{p}$ and range $\hat{r}$, corresponding to the target, and invert~\eqref{eq:RxPower} to get
\begin{equation}
\hat{\Psi} = \hat{p} \frac{(4\pi)^3 \hat{r}^4 f_c^2}{P_T N M G_T G_R c_0^2} \; .
\end{equation}

\begin{remark}
The values in Table \ref{tab:max_sense_ranges} suggest that typical objects of interest can be sensed kilometers away. However, those are the maximum achievable ranges due to the noise limit, determined with \eqref{eq:AchievableRangeNoise}. As discussed in Section~\ref{sec:model}, quantization noise,  resolution limitations, and unambiguous range must be considered as well. In what follows, we show that the last mentioned constraints are the dominating factors in practice in all considered use cases. Therefore, one could deduce that sensing performance is - for the systems we assumed in Section~\ref{sec:systemParameters} -  typically not limited by thermal noise.
\end{remark}

{\renewcommand{\arraystretch}{1.2}
\begin{table}[hbt!]
\centering
\caption{Max. range for different objects due to thermal noise limit (SNR = $\gamma^*$ = 17 \MakeTextLowercase{d}B)}
\label{tab:max_sense_ranges}
\resizebox{\columnwidth}{!}{%
\begin{tabular}{|l?c|c|c|c|}
\hline
\multicolumn{5}{|c|}{\textbf{Outdoor ($P_{T, O}$)}} \\
\Xhline{2.5\arrayrulewidth}
\textbf{Object} & \textbf{$\Psi$ [m\textsuperscript{2}]} & $r^*_n$ \textbf{(\gls{fr1}) [km]} & $r^*_n$ \textbf{(\gls{fr2}) [km]} & $r^*_n$ \textbf{(\gls{fr3}) [km]} \\
\Xhline{2.5\arrayrulewidth}
Drone & 0.1 & 7.65 & 3.01 & 12.23 \\ \hline
Human & 1 & 13.39 & 5.27 & 21.44 \\ \hline
Car & 100 & 41.18 & 16.51 & 65.99 \\ 
\Xhline{2.5\arrayrulewidth}
\multicolumn{5}{|c|}{\textbf{Indoor ($P_{T, I}$)}} \\
\Xhline{2.5\arrayrulewidth}
Drone & 0.1 & 2.99 & 1.63 & 3.19 \\ \hline
Human & 1 & 5.23 & 2.85 & 5.58 \\ \hline
AGV & 2 &  6.19 & 3.37 & 6.61 \\ \hline

\end{tabular}
}
\end{table}
}
\begin{remark}
The precise extent of the quantization impact is highly dependent on the scenario. Since the presence of strong close reflectors is hard to model, we will for the most part disregard the quantization limitation in the following  use case analysis. However, it must be kept in mind when evaluating/planning specific sensing deployments.
\end{remark}
\begin{remark}
The authors would like to remark that the proposed analysis in this paper allows to initially determine the feasibility of a use case. For more precise performance forecasts, one should refer to ray tracing and corresponding simulations based on signal propagation~\cite{arnold2022maxray,bauhofer2023multi}, with increased implementation effort and loss of generality. A further -- needed -- step towards assessment is the validation with real measurements~\cite{wild2023integrated,nuss2023frequency}, with a drastic increase in invested time and resources.  
\end{remark}


{\renewcommand{\arraystretch}{1.5}
\begin{table*}[hbt!]
\centering
\caption{Summary of investigated use cases}
\label{tab:use_cases_overview}
\begin{tabular}{|l?l|l|l|}

\hline
 \textbf{Use Case} & \textbf{Frequency Range(s)} & \textbf{Limiting Factor(s)} & \textbf{Max. Sensing Range $r^*$}  \\ 
\Xhline{3\arrayrulewidth}
Indoor Factory Safety & FR2 & \makecell[l]{ NLOS conditions, \\ System resolution} & \makecell[l]{15 m for $\rho_h^* = 0.5$ m, \\ probably lower due to NLOS} \\ 
\Xhline{3\arrayrulewidth}
Traffic Count & FR2, FR3 & Horizontal resolution & 50 -- 100 m  \\ 
\hline
Ghost Driver Detection & FR2, FR3 & Horizontal resolution & 100 -- 200 m  \\ 
\hline
Pedestrian Crossing Detection & FR2, FR3 & Horizontal resolution & 20 -- 40 m \\ 
\Xhline{3\arrayrulewidth}

Outside Drone Detection & FR1, FR3, FR2 (less preferred) & \makecell[l]{Unambiguous range, \\ NLOS conditions}  & \makecell[l]{5000 m (FR1), \\ 2500 m (FR3), \\ 1250 m (FR2)}  \\ 
\hline
\end{tabular}
\end{table*}
}

\subsection{Indoor Factory Safety}\label{sec:indoor_fac}

As the first use case, an indoor factory scenario is considered. Such a typically cluttered environment with multiple and possibly closely spaced reflectors requires high spatial resolution capabilities, achievable via a high total bandwidth $B$ (for range resolution $\rho_r$) and a large number of antennas (for angular resolution $\rho_{\theta}$ and $\rho_{\phi}$). 
Moreover, a good velocity resolution $\rho_s$ provides another degree of freedom, that may be leveraged to discriminate moving targets of interest from reflections caused by static background clutter. 
We therefore regard systems in \gls{fr3} and especially \gls{fr2} as suitable for this scenario, since the larger path loss in the millimeter-wave range can be tolerated as only short distances, \iec $r\approx10$-$20$~m,  must be supported inside a factory. 
On the other hand, the achievable resolutions $\rho_r = 0.76$~m and $\rho_h = 0.133r \approx 1.3$-$2.7$~m in \gls{fr1} will presumably not be sufficient to cover most interesting applications, as it becomes harder to separate objects of interest from the background clutter.
\\Sufficient distances will likely be possible for sensing objects of interest, such as humans or \glspl{agv}, even in case the maximum achievable range is considerably reduced by the presence of strong reflecting objects due to quantization.
Therefore, in cluttered factory environments, we think that resolution will be the limiting factor. For instance, if one aims at detecting objects at similar range with a horizontal separation of $\rho_h^* = 0.5$~m, one could achieve this performance up to a range of $r^* = 15.25$ m from the sensing point in \gls{fr2} and \gls{fr3}, while only $r^* = 3.8$~m would be possible for \gls{fr1}. In \gls{fr2}, the high range resolution of $\rho_r = 0.1~$m may additionally be leveraged, which is why we consider \gls{fr2} most suitable for indoor factories.
Nonetheless, due to blockers and \gls{nlos} conditions, multiple \glspl{trp} will likely still be required to offer reliable sensing services over a wide area in a factory. 




\begin{remark}
Since we always assume an equal \gls{snr} $\gamma^* = 17$ dB -- and not an equal sensing range -- for determining the sensing \glspl{kpi} in Table~\ref{tab:fr1_fr2_perf_params}, better accuracies can obviously be achieved for sensing a stronger reflector (\gls{agv} in this case) at the same distance as a weaker reflector (human body), due to the larger \gls{rcs} and thus resulting higher \gls{snr}.
\end{remark}

\subsection{Roadway Monitoring}\label{sec:highway_mon}

The second family of use cases in this document is concerned with roadway monitoring. 
In that context, applications such as traffic count, ghost driver detection (\iec detecting a vehicle moving in the wrong direction of traffic on a highway), or pedestrian/animal detection are conceivable and will be addressed in the following. 
\\ To cover as much of the street as possible, it may initially appear favorable to operate in \gls{fr1}. Most use cases, however, will not have to monitor the entire roadway (especially in case of long highways), but only certain points of interest (\egc traffic count) or at regular intervals (\egc ghost driver detection). Moreover, such applications will likely be limited by the resolution rather than by the range, so in the following we mainly consider \gls{fr2} and \gls{fr3}.

\subsubsection{Traffic Count}
In view of this use case, the spatial resolution capabilities are of interest, since for traffic count it is required to discriminate two vehicles driving in different lanes. As introduced in Section~\ref{sec:model}, horizontal and vertical direction resolution are considered separately. 
We further assume that the base station ``observing" the roadway is placed at a height of a few meters, such that vehicles driving within a lane can be treated almost as \textcolor{black}{radial} movements, \iec towards the base station. To discriminate vehicles in different lanes, the horizontal direction resolution of $\rho_h=0.032r$ from Table~\ref{tab:fr1_fr2_perf_params} can then be leveraged. 
This allows distinguishing two vehicles on separate lanes up to a distance of ca. $r^*=78$~m, where a spacing of at least $\rho_h^* = 2.5$~m between them is assumed. While this value represents the horizontal resolution at boresight, which is not always attainable in practice, the high range and speed resolutions can further help. As in the factory scenario, one can make use of the range resolution of ca. 0.1~m in \gls{fr2}, which even allows discriminating different vehicles in the same lane. Moreover, for counting traffic it is enough to merely detect the presence of multiple vehicles rather than being able to perfectly resolve them. 

\begin{remark}
The ability to resolve different objects also depends on the objects' dimensions, as \egc a long truck might cause several reflections. Those influences are again hard to model, however, and are therefore omitted from the discussions in this work.
\end{remark}

\subsubsection{Ghost Driver Detection} Ghost Driver Detection is in many regards similar to the traffic count use case. The requirements are now slightly less stringent, since a ghost driver moves in the opposite direction than cars on the same side of the road such that the Doppler (\iec speed/velocity) domain can be used to discriminate them. Nonetheless, a ghost driver must still be distinguishable from other cars driving in the same direction at similar velocities, but using the correct lane.
Accordingly, horizontal resolution is necessary also in this case. Two cars on opposite lanes could be distinguished up to a distance of ca. $r^*=156$~m, where a spacing of at least $\rho_h^* = 5$~m between them is now assumed due to the higher spacing of lanes in opposite directions. It should be noted that tracking techniques can further help discriminating different vehicles from the history of collected measures, likely further extending the achievable range in practice.

\subsubsection{Pedestrian Crossing Detection}

As a consequence of the previously discussed resolution limitations, it appears unfeasible to reliably monitor big segments of a roadway for people (or animal) crossing detection use cases. 
However, sensing systems may still be deployed  to detect humans in particularly dangerous areas, \egc at railway crossings. The spatial resolution capabilities will likely again be the limiting factor, as pedestrians on a crossing should be distinguishable from people moving along ``safe paths" next to the road or railway. Hence, \gls{fr2} and \gls{fr3} are again preferable due to offering the best spatial resolution capabilities. Assuming a required $\rho_h^* = 1$~m leads to an achievable range of $r^*=31.25$~m. 

\subsection{Outside Drone Detection}\label{sec:drone_detection}
As the last use case, outside drone detection is discussed. Since drone detection may be of interest in different settings/environments (\egc urban macro, airport, etc.), all three frequency ranges can be considered for this application. 
Operating in \gls{fr1} enables a higher achievable sensing range and allows to leverage deployments designed mainly for communications coverage, but also offers worse capabilities \wrt accuracy and resolution. This trade-off in favor of the sensing range, however, appears acceptable, as typically drones fly in open space, and detection is more important than precisely localizing and counting them, for which resolution plays a critical role. 
According to \eqref{eq:AchievableRangeNoResolution} and system parameters, the system will be limited mainly by unambiguous range, resulting in $r^* = 5000$~m for \gls{fr1}, $r^* = 1250$~m in \gls{fr2}, and $r^* = 2500$~m for \gls{fr3}. Therefore, operating in \gls{fr1} or \gls{fr3} is generally advocated for the use case of outside drone detection. \gls{fr2} should also not be precluded, especially if the scenario requires higher resolution capabilities and/or the maximum required sensing range is low.

Finally, Table~\ref{tab:use_cases_overview} summarizes this investigation by providing an overview of the suggested frequency range(s), limiting factor(s), and maximum achievable sensing range for each use case.

%% file: Content/Conclusion.tex
\section{Conclusion}\label{sec:conclusion}

In this paper, we have derived an analytical model for \gls{isac} performance assessment taking practical limitations into account. Based on our framework and system paremeterizations that can be expected for the main \gls{6G} frequency ranges, we evaluated emerging \gls{isac} use cases \wrt sensing \glspl{kpi}, and drew first preliminary conclusions about their feasibility. Further, visual examples of the most meaningful practical limitations of \gls{isac} systems have been provided by means of simulations.

We concluded that the main limiting factor for sensing in \gls{6G} \gls{isac} systems will be \gls{los} coverage and spatial resolution. 
While the first can be addressed with densification accompanied by considerable costs, the second can be addressed by further increasing array sizes of current \gls{isac} systems, especially in the horizontal direction.

While this work only represents a first step towards a complete analytical characterization of \gls{isac} systems, the authors believe that it can help in better understanding their practical limitations together with their implications on possible use cases.

%% file: Content/Appendices.tex

\begin{table}[ht]%
\centering
\caption{Notation used in this document}
\label{tab:Notation}
\begin{tabular}{|c|c|}
\hline
$P_R$ & Received power [W]\\ \hline
$P_T$  & Transmitted power [W] \\ \hline
$P_{T, O}$ & Transmitted power in outdoor scenarios [W] \\ \hline
$P_{T, I}$ & Transmitted power in indoor scenarios [W] \\ \hline
$P_N$ & Noise power [W] \\ \hline
$N_0$ & Thermal noise spectral density = -174 [dBm/Hz] \\ \hline
$F$ & Noise figure \\ \hline
$\gamma_S$ & \gls{snr} on each \gls{ofdm} sample \\ \hline
$\gamma$ & \gls{snr} after processing gains \\ \hline
$\gamma^*$ & Min. required \gls{snr} \\ \hline
$\gamma_q$ & Max. \acrshort{sqnr} at the receiver \\ \hline
$G_T$ & Total transmit antenna array gain  \\ \hline
$G_R$ & Total receive antenna array gain  \\ \hline
$G_E$ & Single array element gain  \\ \hline
$S_0$ & Maximum transmit power density [$10$ W/m\textsuperscript{2}] \\ \hline
$T^*$ & \gls{tdd} duty-cycle \\ \hline
$P^*$ & \begin{tabular}{@{}c@{}}Electromagnetic field limit power reduction \\ factor due to beam steering and time averaging \end{tabular}  \\ \hline
$r$ & Range [m] \\ \hline
$r^*$ & Max. supported range [m] \\ \hline
$\alpha$ & Isolation between transmitter and receiver  \\ \hline
$\Psi$ & Radar cross section [m\textsuperscript{2}] \\ \hline
$\lambda$ & Wavelength [m] \\ \hline
$c_0$ & Speed of light in vacuum [m/s]\\ \hline
$f_c$ & Central frequency [Hz]\\ \hline
$\Delta f$ & Subcarrier spacing [Hz]\\ \hline
$N$ & Number of subcarriers \\ \hline
$T_0$ & \gls{ofdm} symbol duration incl. cyclic prefix [s] \\ \hline
$T_f$ & Frame duration [s] \\ \hline
$M$ & Number of \gls{ofdm} symbols \\ \hline
$R, C$ & Number of antenna rows, columns \\ \hline
$\Delta r, \Delta c$ & Antenna spacing between rows, columns [m] \\ \hline
$\sigma_r$ & Range standard deviation [m] \\ \hline
$\sigma_s$ & Speed standard deviation [m/s] \\ \hline
$\sigma_z$ & \begin{tabular}{@{}c@{}} Standard deviation \wrt vertical array \\ direction (z-axis)
[NAF] \end{tabular}  \\  \hline
$\sigma_{\phi}$ & Elevation standard deviation [$^\circ$] \\ \hline
$\sigma_x$ & \begin{tabular}{@{}c@{}} Standard deviation \wrt horizontal array \\ direction (x-axis)
[NAF] \end{tabular}  \\  \hline
$\sigma_{\theta}$ & Azimuth standard deviation [$^\circ$] \\ \hline
$\sigma_t$ & Clock timing error standard deviation [s] \\ \hline
$\sigma_f$ & Clock frequency error standard deviation [Hz] \\ \hline
$\rho_r$ & Range resolution [m] \\ \hline
$\rho_s$ & Speed resolution [m/s] \\ \hline
$\rho_z$ &\ \begin{tabular}{@{}c@{}} Resolution \wrt vertical array \\ direction (z-axis)
[NAF] \end{tabular} \\ \hline
$\rho_{\phi}$ & Elevation resolution [$^\circ$] \\ \hline
$\rho_v$ & Vertical direction resolution [m] \\ \hline
$\rho_x$ &\ \begin{tabular}{@{}c@{}} Resolution \wrt horizonal array \\ direction (x-axis)
[NAF] \end{tabular} \\ \hline
$\rho_{\theta}$ & Azimuth resolution [$^\circ$] \\ \hline
$\rho_h$ & Horizontal direction resolution [m] \\ \hline
$r_u$ & Unambiguous range [m] \\ \hline
$s_u$ & Unambiguous speed [m/s] \\ \hline

\end{tabular}
\end{table}

%% file: Content/Acknowledgment.tex
\section*{Acknowledgments}
\if\paperversion1
\input{Content/Acknowledgment_External}
\else
\input{Content/Acknowledgment_Internal}
\fi

%% file: Content/Acknowledgment_External.tex
The authors would like to thank Christophe Grangeat, Harish Viswanathan, Frank Schaich, Artjom Grudnitsky, Junqing Guan, and Mark Doll for their insights and feedback during the paper's writing.

This work was developed within the KOMSENS-6G project, partly funded by the German Ministry of Education and Research under grant 16KISK112K.

%% file: use_cases_main.bbl
\begin{thebibliography}{10}
\providecommand{\url}[1]{#1}
\csname url@samestyle\endcsname
\providecommand{\newblock}{\relax}
\providecommand{\bibinfo}[2]{#2}
\providecommand{\BIBentrySTDinterwordspacing}{\spaceskip=0pt\relax}
\providecommand{\BIBentryALTinterwordstretchfactor}{4}
\providecommand{\BIBentryALTinterwordspacing}{\spaceskip=\fontdimen2\font plus
\BIBentryALTinterwordstretchfactor\fontdimen3\font minus
  \fontdimen4\font\relax}
\providecommand{\BIBforeignlanguage}[2]{{%
\expandafter\ifx\csname l@#1\endcsname\relax
\typeout{** WARNING: IEEEtran.bst: No hyphenation pattern has been}%
\typeout{** loaded for the language `#1'. Using the pattern for}%
\typeout{** the default language instead.}%
\else
\language=\csname l@#1\endcsname
\fi
#2}}
\providecommand{\BIBdecl}{\relax}
\BIBdecl

\bibitem{viswanathan2020communications}
H.~Viswanathan and P.~E. Mogensen, ``Communications in the {6G} era,''
  \emph{IEEE Access}, vol.~8, pp. 57\,063--57\,074, 2020.

\bibitem{liu2022integrated}
F.~Liu \emph{et~al.}, ``{Integrated sensing and communications: Towards
  dual-functional wireless networks for 6G and beyond},'' \emph{IEEE journal on
  selected areas in communications}, 2022.

\bibitem{wild2023integrated}
\BIBentryALTinterwordspacing
T.~Wild, A.~Grudnitsky, S.~Mandelli, M.~Henninger, J.~Guan, and F.~Schaich,
  ``{6G Integrated Sensing and Communication: From Vision to Realization},''
  2023. [Online]. Available: \url{https://arxiv.org/abs/2305.01978}
\BIBentrySTDinterwordspacing

\bibitem{3gpp_38301_1}
3GPP, ``{User Equipment (UE) radio transmission and reception; Part 1: Range 1
  Standalone},'' Technical Specification Group RAN, Technical Specification
  38.301-1, 2023, {Version 17.8.0 (Rel. 17)}.

\bibitem{3gpp_38301_2}
------, ``{User Equipment (UE) radio transmission and reception; Part 2: Range
  2 Standalone},'' Technical Specification Group RAN, Technical Specification
  38.301-2, 2023, {Version 17.8.0 (Rel. 17)}.

\bibitem{henninger2022probabilistic}
M.~Henninger \emph{et~al.}, ``{Probabilistic 5G Indoor Positioning Proof of
  Concept with Outlier Rejection},'' in \emph{2022 Joint European Conference on
  Networks and Communications \& 6G Summit (EuCNC/6G Summit)}.\hskip 1em plus
  0.5em minus 0.4em\relax IEEE, 2022, pp. 249--254.

\bibitem{liu2022survey}
A.~Liu \emph{et~al.}, ``A survey on fundamental limits of integrated sensing
  and communication,'' \emph{IEEE Communications Surveys \& Tutorials},
  vol.~24, no.~2, pp. 994--1034, 2022.

\bibitem{hexa_x}
``{HEXA-X-II project},'' \url{https://hexa-x-ii.eu/}.

\bibitem{braun2014ofdm}
K.~M. Braun, ``{OFDM radar algorithms in mobile communication networks},''
  Ph.D. dissertation, Karlsruher Institut f{\"u}r Technologie (KIT), 2014.

\bibitem{MUSIC_collapse}
B.~A. Johnson, Y.~I. Abramovich, and X.~Mestre, ``{MUSIC, G-MUSIC, and
  Maximum-Likelihood Performance Breakdown},'' \emph{IEEE Transactions on
  Signal Processing}, vol.~56, no.~8, pp. 3944--3958, 2008.

\bibitem{nwankwo2017survey}
C.~D. Nwankwo, L.~Zhang, A.~Quddus, M.~A. Imran, and R.~Tafazolli, ``A survey
  of self-interference management techniques for single frequency full duplex
  systems,'' \emph{IEEE Access}, vol.~6, pp. 30\,242--30\,268, 2017.

\bibitem{bennett1948spectra}
W.~R. Bennett, ``Spectra of quantized signals,'' \emph{The Bell System
  Technical Journal}, vol.~27, no.~3, pp. 446--472, 1948.

\bibitem{behravan2002papr}
A.~Behravan and T.~Eriksson, ``{PAPR and other measures for OFDM systems with
  nonlinearity},'' in \emph{The 5th International Symposium on Wireless
  Personal Multimedia Communications}, vol.~1.\hskip 1em plus 0.5em minus
  0.4em\relax IEEE, 2002, pp. 149--153.

\bibitem{mandelli2022sampling}
S.~Mandelli, M.~Henninger, and J.~Du, ``{Sampling and Reconstructing Angular
  Domains With Uniform Arrays},'' \emph{IEEE Transactions on Wireless
  Communications}, vol.~22, no.~6, pp. 3628--3642, 2023.

\bibitem{hoctor1990unifying}
R.~T. Hoctor and S.~A. Kassam, ``{The unifying role of the coarray in aperture
  synthesis for coherent and incoherent imaging},'' \emph{Proceedings of the
  IEEE}, vol.~78, no.~4, pp. 735--752, Apr. 1990.

\bibitem{wesemann2023energy}
S.~Wesemann, J.~Du, and H.~Viswanathan, ``{Energy Efficient Design of Extreme
  Massive MIMO},'' \emph{arXiv preprint arXiv:2301.01119}, 2023.

\bibitem{holma2021extreme}
H.~Holma, H.~Viswanathan, and P.~Mogensen, ``{Extreme massive MIMO for macro
  cell capacity boost in 5G-advanced and 6G},'' in \emph{White paper}.\hskip
  1em plus 0.5em minus 0.4em\relax Nokia Bell Labs, 2021.

\bibitem{recommendation1999limitation}
{Council of Recommendation}, ``{Limitation of exposure of the general public to
  electromagnetic fields (0 Hz to 300 GHz)},'' \emph{Official Journal of the
  European Communities}, vol. 199, 1999.

\bibitem{baracca2018statistical}
P.~Baracca, A.~Weber, T.~Wild, and C.~Grangeat, ``{A statistical approach for
  RF exposure compliance boundary assessment in massive MIMO systems},'' in
  \emph{WSA 2018; 22nd International ITG Workshop on Smart Antennas}.\hskip 1em
  plus 0.5em minus 0.4em\relax VDE, 2018, pp. 1--6.

\bibitem{3gpp_38211}
3GPP, ``{Physical channels and modulation},'' Technical Specification Group
  RAN, Technical Specification 38.211, 2023, {Version 17.4.0 (Rel. 17)}.

\bibitem{skolnik1980introduction}
M.~I. Skolnik, ``Introduction to radar systems,'' \emph{New York}, 1980.

\bibitem{de2019drone}
{\'A}.~D. de~Quevedo, F.~I. Urzaiz, J.~G. Menoyo, and A.~A. L{\'o}pez, ``{Drone
  detection and radar-cross-section measurements by RAD-DAR},'' \emph{IET
  Radar, Sonar \& Navigation}, vol.~13, no.~9, pp. 1437--1447, 2019.

\bibitem{arnold2022maxray}
M.~Arnold \emph{et~al.}, ``{MaxRay: A raytracing-based integrated sensing and
  communication framework},'' in \emph{2022 2nd IEEE International Symposium on
  Joint Communications \& Sensing (JC\&S)}.\hskip 1em plus 0.5em minus
  0.4em\relax IEEE, 2022, pp. 1--7.

\bibitem{bauhofer2023multi}
\BIBentryALTinterwordspacing
M.~Bauhofer, S.~Mandelli, M.~Henninger, T.~Wild, and S.~ten Brink,
  ``{Multi-Target Localization in Multi-Static Integrated Sensing and
  Communication Deployments},'' 2023. [Online]. Available:
  \url{https://arxiv.org/abs/2306.07740}
\BIBentrySTDinterwordspacing

\bibitem{nuss2023frequency}
B.~Nuss, L.~G. de~Oliveira, and T.~Zwick, ``{Frequency comb MIMO OFDM radar
  demonstrator with high unambiguous velocity},'' \emph{International Journal
  of Microwave and Wireless Technologies}, pp. 1--9, 2023.

\end{thebibliography}
